\documentclass[%
 aip,
 amsmath,amssymb,
 preprint,%
]{revtex4-1}

\usepackage{graphicx}
\usepackage{dcolumn}
\usepackage{bm}
\usepackage{subfigure}
\usepackage[utf8]{inputenc}
\usepackage[T1]{fontenc}
\usepackage{epstopdf}
\usepackage{amsthm,amsmath,amssymb,amsfonts}
\usepackage{mathrsfs}
\usepackage{ulem}
\usepackage{amssymb}
\usepackage{listings}
\usepackage{amsmath}
\usepackage{color}
\usepackage{mathtools}
\usepackage{graphics}
\usepackage{caption}
\usepackage{listings}
\usepackage{tabularx}
\usepackage{appendix}

\begin{document}

\preprint{AIP/123-QED}

\title{Gyrokinetic simulation of the spontaneous toroidal rotation of plasma in a stochastic magnetic field}

\author{Jinxiang You}

\affiliation{%
Department of Engineering and Applied Physics, University of Science and Technology of China, Hefei, 230026, China
}%

\author{Shaojie Wang}
\email{wangsj@ustc.edu.cn}
\affiliation{%
Department of Engineering and Applied Physics, University of Science and Technology of China, Hefei, 230026, China
}%

\date{\today}

\begin{abstract}
     Since the DIII-D resonant magnetic perturbation experiment [Nucl. Fusion $\bm{59}$, 126010 (2019)] suggests that the neoclassical toroidal viscosity due to the collisional effects associated with the non-resonant magnetic perturbations is not enough to explain the observed toroidal rotation, it is of interest to investigate the toroidal rotation induced by the anomalous diffusion due to the resonant magnetic perturbations. Gyrokinetic simulation of the toroidal rotation of plasma in a stochastic magnetic field is carried out to investigate the resonant magnetic perturbations effects on toroidal rotation. The simulation results suggest that, in a stochastic magnetic field, resonant magnetic perturbations drive the plasma to toroidally rotate through the ambipolar radial electric field. It is found that this spontaneous flow driven on the time scale less than an ion-ion collision time is the parallel return flow of the $\bm{E}_r\times\bm{B}_0$ drift, which is  due to the the ambipolar radial electric field induced by the non-ambipolar radial diffusion in the stochastic magnetic field. Collisional effect changes the plasma toroidal rotation from the return flow to the rigid-body flow after a few ion-ion collision times. The toroidal rotation observed in DIII-D resonant magnetic perturbation experiment [Nucl. Fusion $\bm{59}$, 126010 (2019)], can be explained by the rigid-body rotation driven by the ambipolar radial electric field generated by the stochastic magnetic field layer.
\end{abstract}

\maketitle

\section{Introduction}

    The toroidal rotation is a common phenomenon in tokamak experiments, such as JET\cite{JET1997}, Alcator C-Mod\cite{C-Mod2004}, Tore Supra\cite{Supra2001}, DIII-D\cite{DIII-D2007,Schmitz_2019}, and KSTAR\cite{KSTART2012,KSTART2019}. The plasma toroidal rotation can be driven by the injection of neutral beam, or spontaneously generated without external input of toroidal momentum. The driving mechanism of spontaneous toroidal rotation is not well understood. One of the methods to change the toroidal rotation is to apply external magnetic perturbations. The toroidal rotation has important effects on the tokamak plasma confinement, such as, the control of the neoclassical tearing modes (NTMs)\cite{NTM2012} and the suppression of reststive wall modes (RWMs)\cite{D3DRWM2002,Zheng2005,RWM2015,RWM2022}. Understanding the mechanism\cite{Callen2010} and process of spontaneous toroidal rotation in a perturbed magnetic field, we can modify the plasma toroidal rotation by controlling the externally applied magnetic field perturbation, thus improving the confinement performance of tokamak.

    In recent years, the effect of neoclassical toroidal viscosity(NTV) on plasma toroidal rotation, which is caused by the collisional effect associated with the non-resonant components of three-dimensional magnetic perturbations, has been intensively investigated. Researchers have carried out extensive experimental\cite{NSTX2004,NSTX2006,JET2010,KSTART2012,KSTART2013,KSTART2019}, theoretical\cite{Shaing2003,Shaing2008,Shaing2009a,Shaing2009b,Shaing2009c,Shaing2009d,JK-Park2009,JK-Park2011}, and simulation\cite{Sun2010,Sun2011,Sun2013,Liu2013,Sun2019} investigation related to the NTV effect. K. C. Shaing theoretically solved the linear drift kinetic equation and obtained the NTV torques in different collision regimes.\cite{Shaing2003,Shaing2008,Shaing2009a,Shaing2009b,Shaing2009c,Shaing2009d} J. -K. Park et al. gave the analytical expression of NTV torque by using the Krook collision model to solve the linear drift kinetic equation.\cite{JK-Park2009} The explicit relationship between the plasma perturbed kinetic energy and the NTV torque is given in Ref. \onlinecite{JK-Park2011}. Sun et al. developed an NTV torque code NTVTOK\cite{Sun2010,Sun2011,Sun2013} by solving the linear drift kinetic equation numerically, and successfully applied it in the experimental analysis\cite{Sun2019,EAST2019} of tokamak experiments. The NTVTOK was also successfully coupled with the MHD code MARS-F\cite{Liu2010,Liu2013}.

    The essence of NTV torque is a collisional transport phenomenon. The non-resonant component of the perturbed magnetic field affects the banana orbit of the trapped particles, generating additional radial drift, resulting in enhanced neoclassical transport. The different neoclassical transport coefficients of ions and electrons result in charge separation and generate a non-ambipolar radial electric field. This radial electric field can be equivalent to a toroidal torque, that is, the NTV torque. In a stochastic magnetic field, the magnetic field line braiding results in anomalous radial diffusion of a collisionless plasma, with the electron diffusivity much larger than the ion diffusivity\cite{RR1978}; this non-ambipolar collisionless diffusion leads to the charge separation, and an ambipolar radial electric field will be established quickly\cite{Harvey1981,jxyou2023}. Similarly, this ambipolar radial electric field  will also drive the plasma to toroidally rotate. In the DIII-D experiment\cite{Schmitz_2019}, when a stochastic magnetic field is generated at the edge of the device by using an external resonant magnetic perturbations(RMPs), the radial electric field well in the edge plasma is reversed, and the toroidal rotation velocity increases significantly in the direction of plasma current. This RMP experiment\cite{Schmitz_2019} suggests that the NTV due to the collisional effects associated with the non-resonant perturbation is not enough to explain the observed toroidal rotation. Therefore, it is of interest to investigated the toroidal rotation induced by the collisionless anomalous diffusion due to the RMPs.

   In this work, the gyrokinetic code NLT\cite{Ye2016,Xu2017} is used to investigate the mechanism and process of plasma toroidal rotation driven by RMPs through an ambipolar radial electric field in a stochastic magnetic field. The remainder of this paper is organized as follows. In Sec. \uppercase\expandafter{\romannumeral2}, fundamental equations used in simulation of spontaneous toroidal rotation in a stochastic magnetic field will be briefly introduced. In Sec. \uppercase\expandafter{\romannumeral3}, the simulation results will be presented and discussed. In Sec. \uppercase\expandafter{\romannumeral4}, the summary is presented.


\section{Fundamental equations}

\subsection{Gyrokinetic Vlasov equation}

In this subsection, we briefly review the gyrokinetic theory, which shall be used in this paper. The vector potential of magnetic perturbations is given by $\delta \boldsymbol{A} = \delta A_{\|} \boldsymbol{b}_0$, $\boldsymbol{b}_0 = \frac{\boldsymbol{B}_0}{B_0}$, with $\boldsymbol{B}_0$ the equilibrium magnetic field. The radial electric field is given by $ \delta \boldsymbol{E}_r = - \nabla \delta \phi(r)$, with $\delta \phi$ the electrostatic potential perturbation, $r$ the minor radius. The dynamic of particles is described by the collision gyrokinetic Vlasov equation with gyrokinetic ions and drift-kinetic electrons. In the gyrocenter coordinates $\boldsymbol{Z} = (\boldsymbol{X},v_{\|},\mu)$, the gyrokinetic Vlasov equation is written as 
\begin{align}
  \partial_t f_s + \dot{\boldsymbol{X}} \partial_{\boldsymbol{X}} f_s + \dot{v}_{\|} \partial_{v_{\|}} f_s = \left( \frac{\partial f_s}{\partial t} \right)_c.
\end{align}
Here,
\begin{subequations}
 \begin{align}
 & \dot{\boldsymbol{Z}} = \{\boldsymbol{Z},H_0 + H_1 \}_0, \label{eq.orbit}  \\
 & H_0 = \frac12 M_s v_{\|}^2 + \mu B_0,  \label{eq.drift}  \\
 & H_1 = e_s \left\langle \delta \phi \right\rangle_{GA} - e_s v_{\|}\left\langle \delta A_{\|} \right\rangle_{GA},
 \end{align}
\end{subequations}
$\left\langle \cdot \right\rangle_{GA}$ is the gyro-average operator defined by 
\begin{align} 
  \left\langle \delta \phi \right\rangle_{GA}(\boldsymbol{X},\mu) \equiv \frac{1}{2\pi} \int_{0}^{2\pi} \delta \phi \left( \boldsymbol{X} + \boldsymbol{\rho} ( \mu,\xi ) \right) d\xi.
\end{align}
Where $f_s$ is gyrocenter distribution function, $\boldsymbol{X}$ is the position of gyrocenter, $v_{\|}$ is the parallel velocity of the gyrocenter, $\mu$ is the magnetic moment, $\xi$ is the gyro-angle, $e_s$ is the particle charge, $M_s$ is the particle mass, the subscript $s$ denote the particle species, $\{ \cdot \}_0$ is the unperturbed Poisson bracket. The gyrokinetic Vlasov equation is solved by using the I-transform method.\cite{Wang12,Wang13}

   Collisions are calculated by using the Krook collision operater,
\begin{align}
\left( \frac{\partial f}{\partial t} \right)_c = - \frac{1}{\tau_c} \left[ f_s-f_{sM,eff}(N_s,T_s,U_s) \right].
\end{align}
Here, $f_{sM,eff}(N_s,T_s,U_s)$ is the equivalent Maxwellian distribution, which has the particle density $N_s(\bm{X})$, temperature $T_s(\bm{X})$, parallel fluid velocity $U_s(\bm{X})$, same as the real distribution $f_s$. Note that this generalized Krook collision term conserves the particle number, energy, and momentum.

In this paper, we only consider the ion-ion collision. The ion-ion collision time is
\begin{align}
     \tau_c = \tau_{ii} = 12 \pi^{3/2} \frac{\varepsilon_0^2 M_i^{1/2} T_i^{3/2}}{N_{0i} e_i^4 \ln \Lambda},
\end{align}
where $\ln \Lambda$ is the Coulomb logarithm.  

\subsection{Gyrokinetic quasi-neutrality equation}

The electrostatic potential is calculated from the gyrokinetic quasi-neutrality equation\cite{WWLEE1983}, which is written as 
\begin{align} \label{quasiNeutrality}
  \nabla \cdot (\frac{N_{0i}M_i}{e_i B_0^2} \nabla_{\perp} \delta \phi ) = \delta N_e - \delta N_i.
\end{align}
The left-hand side of Eq. \eqref{quasiNeutrality} is the ion polarization density, with the long-wavelength approximation used. Here, $\delta N_i$ and $\delta N_e$ represent the perturbation of gyrocenter  density of ion and electron, respectively. 
\begin{align} 
  \delta N_s = \int \left\langle \delta f_s \right\rangle_{GA} d^3 \boldsymbol{v}.
\end{align}
In this work, we consider the passive transport induced by a given three-dimensional magnetic perturbation. For the electrostatic perturbation, we only consider the self-consistently generated radial electrostatic field $\delta\bm{E}_r=-\nabla \delta \phi(r)$ due to the non-ambipolar radial diffusion, therefore, the quasi-neutrality equation solved in practice is given as follows. 
\begin{align} 
  \nabla \cdot (\frac{N_{0i}M_i}{e_i B_0^2} \nabla_{\perp} \delta \phi ) = \left\langle \delta N_e \right\rangle_{FA} - \left\langle \delta N_i \right\rangle_{FA}.
\end{align}
Here, the magnetic surface averaged operator $\left\langle \cdot \right\rangle_{FA}$ is defined as
\begin{align}
   \left\langle  \delta N_s \right\rangle_{FA} \equiv \frac{\int_{0}^{2 \pi} d\zeta \int_{0}^{2 \pi} d\theta J_{\boldsymbol{X}} \delta N_s }{\int_{0}^{2\pi}d\zeta \int_{0}^{2\pi} d\theta J_{\boldsymbol{X}}},
\end{align}
with $J_{\boldsymbol{X}}$ the Jacobian of the real space, $\theta$ the poloidal angle, $\zeta$ the toroidal angle.

Note that this reduced model of radial electrostatic field was used previously in XGC simulation of passive transport in a stochastic magnetic field\cite{Park2010}.

\section{simulation of spontaneous toroidal rotation in a stochastic magnetic field}  \label{sec.rotaSim}

\subsection{simulation set-up}

\begin{figure}[htbp]  
 \includegraphics[scale=0.75]{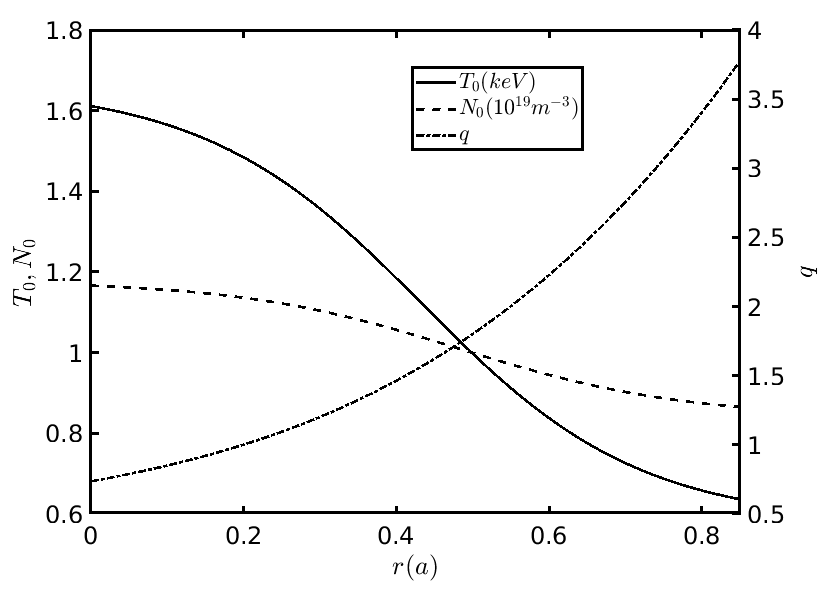}
 \caption{The equilibrium set-up.}
  \label{Equilibrium}
\end{figure}

    In this work, the equilibrium similar to the TEXTOR tokamak is used for simulation set-up. The parameters are major radius $R_0 = 1.75m$, the minor radius $a=0.46m$, and the magnetic field at magnetic axis $B_c = 2.25T$. The profiles of the safety factor $q$, the equilibrium temperature $T_{0i} = T_{0e}=T_0$, the equilibrium density $N_{0i}=N_{0e}=N_0$, are shown in Fig. \ref{Equilibrium}. In order to improve the calculation efficiency, in simulations the electron mass is the realistic mass $M_e = 9.1 \times 10^{-31}kg$, the ion mass $M_i = 100M_e$, and $e_i = -e_e$. The stochastic magnetic field is described by the vector potential $\delta \boldsymbol{A} = \delta A_{\|} \boldsymbol{b}_0$, which is set as
\begin{subequations}
\begin{align} 
&\delta A_{\|} = \sum_{m,n} \delta A_{m,n}\left(r\right) \cos \left( m\theta - n\zeta + \Phi_{m,n} \right),  \\
& \delta A_{m,n}\left(r\right) = A_0 \exp \left[- \left( \frac{r-r_{m,n}}{\sigma_{m,n}} \right)^4 \right].
\end{align}
\end{subequations}
Here, the toroidal mode number $n=20$, the poloidal mode number $38 \leq m \leq 49$, the random phase $\Phi_{m,n}$ is time-independent, $r_{m,n}$ is the minor radius of the rational surface, $q\left(r_{m,n}\right) = m/n$, $\sigma_{m,n} = 0.02 a$, and $A_0 = 3.0 \times 10^{-6}$ $T\cdot m $. The RMPs parameters set in this work are the same as the parameter of Ref.\onlinecite{jxyou2023} to ensure that the magnetic islands overlap sufficiently to produce a completely stochastic magnetic field. The magnetic stochastic layer lies in $ 0.525a \leq r_{m,n} \leq 0.645a$. The number of grid points in the $(r,\alpha,\theta,v_{\|},\mu)$ directions is
$(900,46 \times 20,16,64,16)$. In order to make the radial resolution to achieve the resolution of the magnetic island width, a non-uniform grid is used in the radial direction, and the number of grid points in the stochastic magnetic layer is 300.

   Note that the magnetic perturbation is preprogramed, is not consistently evolved in our simulation; therefore, it is the passive transport induced by magnetic perturbations that is disscussed here, as is similar to Ref.\onlinecite{Park2010}.

 \subsection{spontaneous toroidal flow in a stochastic magnetic field with a self-consistent radial electric field} \label{sec.amEr}

\begin{figure}[htbp] 
  \includegraphics[scale=0.48]{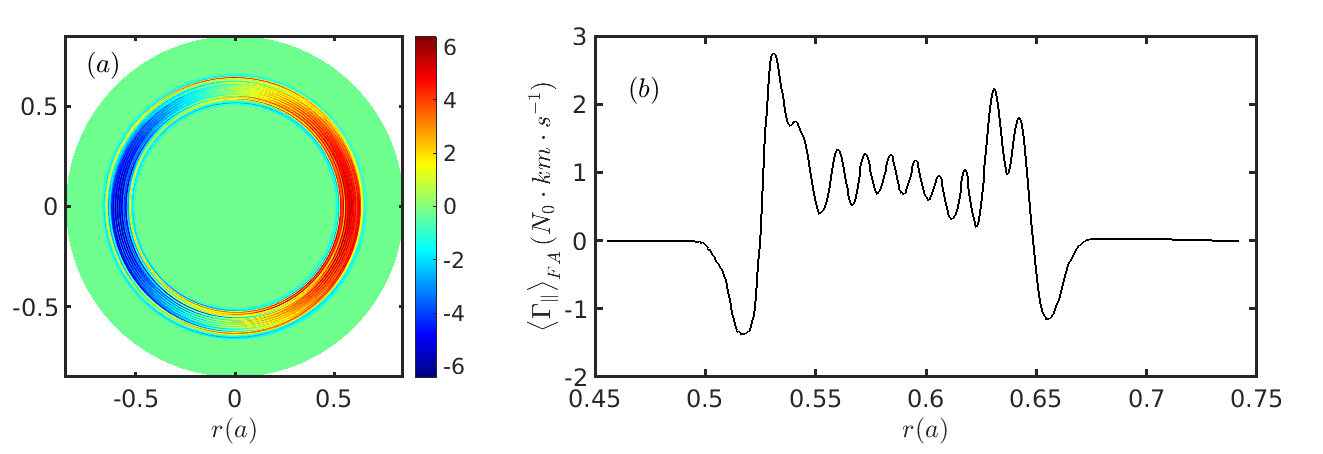} 
\caption{Ion toroidal rotation spontaneously driven in a stochastic magnetic field. (a) $\Gamma_{\|}(r,\theta)$, and (b) $\left\langle  \Gamma_{\|} \right\rangle_{FA}(r)$.}  \label{CosM}
\end{figure}

    In this subsection, collision will not be considered for the time being and simulation results of the passive transport of gyrokinetic ion and drift-kinetic electron in a stochastic magnetic field with a self-consistent radial electric field will be presented. Since the electron mass is much smaller than ion mass, the simulated time step is limited by the characteristic motion time of the electron in the kinetic simulation; the time step in this simulation is 100 electron gyro-period.

The self-consistent radial electric field is obtained by solving the gyrokinetic quasi-neutral equation in Sec.\uppercase\expandafter{\romannumeral2}. In this simulation, the different radial transport coefficients of electrons and ions lead to charge separation, consequently generate a radial electric field, which prevents further charge separation. The particle flux interacts with the radial electric field to form oscillations. The ion and electron particle flux reaches the ambipolarity condition on the time scale($10^{-7}\mbox{-}10^{-6}s$) much less than an ion-ion collision time($10^{-4}\mbox{-}10^{-3}s$). The saturated radial ambipolar electric field is $2.4 kV \cdot m^{-1}$, which is consistent well with the theoretical value. This process of the ambipolar radial electric field formation has been investigated in detail in Ref. \onlinecite{jxyou2023}.

	Since the toroidal velocity $v_{\zeta} \approx v_{\|}$, the ion toroidal flow can be approximated by the parallel flow, which is given by
\begin{align} \label{Eq.para}
   \Gamma_{\zeta} \approx \Gamma_{\|} =  \int v_{\|} \delta f_i d^3\boldsymbol{v}.
\end{align}

	The ion toroidal flow found in this simulation is shown in Fig. \ref{CosM}. Fig. \ref{CosM}(a) indicates that in the stochastic magnetic field layer, the poloidal structure of $\Gamma_{\|}$ is a cosine structure, and the maximum value of $\Gamma_{\|}$ is about $5 N_0 \cdot km\cdot s^{-1}$. The radial structure of $\left\langle  \Gamma_{\|} \right\rangle_{FA}$ is shown in Fig. \ref{CosM}(b).

\subsection{effects of the ambipolar radial electric field on the toroidal rotation in a stochastic magnetic field}  \label{sec.effreturn}

\begin{figure}[htbp] 
  \includegraphics[scale=0.57]{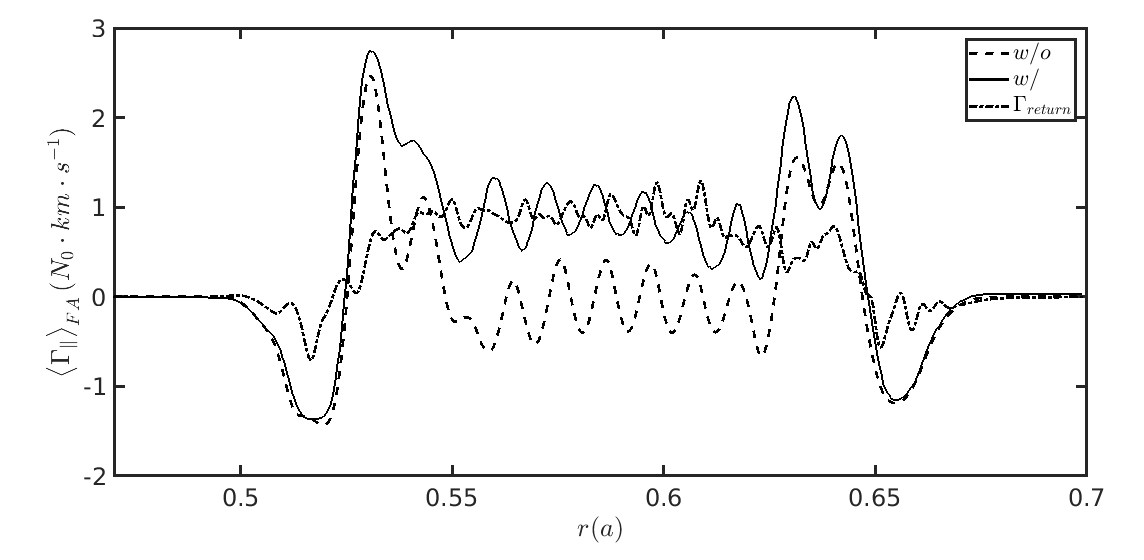}
  \caption{Ion toroidal flow $\left\langle  \Gamma_{\|} \right\rangle_{FA}$ in a stochastic magnetic field. Black line: self-consistent simulation with the ambipolar radial electric field. Dashed line: simulation without the radial electric field. Dot-dashed line: theory.    }
  \label{WoR}
\end{figure}

Perturbations in this simulation system include the given magnetic perturbation and the self-consistently generated ambipolar radial electrostatic field. To understand the driving mechanism of the toroidal flow, we carried out a simulation with the radial electric field switched off, but with the magnetic perturbation retained in the system; the results are shown in Fig. \ref{WoR}. The results indicate that the driven flow is much lowered with the ambipolar radial electric field switched off; in fact, the volume-averaged flow driven in this case is almost zero. This suggests that the toroidal flow driven in a stochastic magnetic field is due to the ambipolar radial electric field. 

The pattern of the flow driven in the self-consistent simulation case suggests that it is a return flow. The familiar Pfirsch-Schl\"uter return flow \cite{wesson2011tokamaks}, which is driven by the diamagnetic drift, is given by
\begin{align}  \label{eq.PS}
   \Gamma^{PS} = 2q \frac{-\partial_r p_i}{e_i B_c} \cos \theta,
\end{align}
with $p_i$ the ion thermal pressure.

To find the return flow driven by the $\delta \bm{E}_r\times\bm{B}_0$ drift, one can simply replace the ion thermal pressure gradient $-\partial_r p_i$ in Eq. \eqref{eq.PS} with $e_i N_{0i} \delta E_r$. 
\begin{align}  \label{eq.return}
   \Gamma_{return} = 2q N_0 \frac{\delta E_r}{B_c} \cos \theta.
\end{align}
Although Eq. \eqref{eq.return} can be derived by using the fluid model, a drift kinetic derivation of Eq. \eqref{eq.return} may be useful for the readers, since it is the kinetic simulation results that we are discussing here. The details of the derivation of Eq. \eqref{eq.return} are shown in Appendix \ref{Appe.PS}. 

Substituting the simulated ambipolar radial electric field into Eq. \eqref{eq.return}, we found the theoretically predicted return flow, which is shown in Fig. \ref{WoR}. The agreement between the theory and simulation indicates that the spontaneous toroidal flow driven in a stachastic magnetic field on the time scale shorter than the ion-ion collision time is indeed the return flow driven by the ambipolar radial electric field. 

\subsection{formation of the rigid-body toroidal rotation} \label{sec.rigid}

  Note that the ambipolar radial electric field is formed on the electron transit time scale in the stochastic magnetic field layer\cite{jxyou2023}. After the ambipolar radial electric field is established, the parallel return flow is driven on the ion transit time scale, which is typically shorter than an ion-ion collision time and discussed in Appendix \ref{Appe.PS-S}. Clearly, it is reasonable to ignore the collisional effects in the simulation of the return flow formation.

On the time scale longer than the ion-ion collision time, collisional effects should be considered. According to Ref. \onlinecite{Hinton1985}, the equilibrium toroidal rotation induced by a strong radial electric field $\bm{E}_r=-\nabla\phi(r)$, which generates a radial force much stronger than the pressure gradient force, on the time scale longer than the ion-ion collision time, is the rigid-body toroidal rotation. 
\begin{equation} \label{eq.rigid}
\bm{\Gamma_{eq}}=N_0(\psi)\frac{-d\phi}{d\psi}R^2\nabla\zeta\approx N_0\frac{E_r}{B_{\theta}}\bm{e}_{\zeta}, 
\end{equation}
with $\psi(r)$ the poloidal magnetic flux, $B_{\theta}$ the poloidal magnetic field. Note that the radial ambipolar electric field discussed here is moderate, which generates a radial force comparable to the pressure gradient force.

   In this subsection, collision will be considered in simulation, and the simulation results of the formation process of the rigid-body rotation will be presented. 

\begin{figure}[htbp] 
  \includegraphics[scale=0.4]{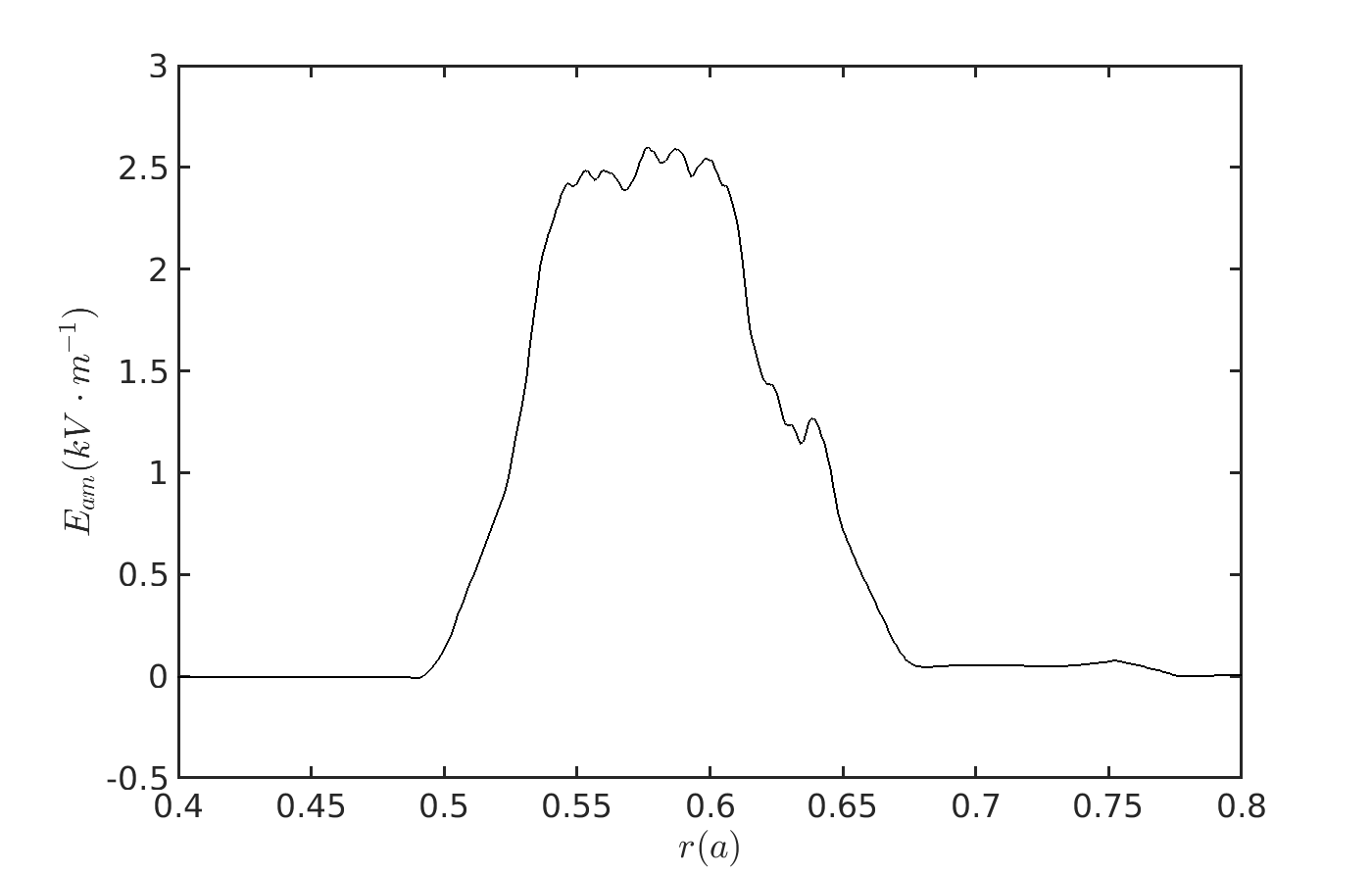}
  \caption{The ambipolar radial electric field formed in stochastic magnetic perturbations on the time scale shorter than an ion-ion collision time.}
  \label{amEr}
\end{figure}

Since we do not consider the collisional NTV effect here, in this subsection, we do not include the magnetic perturbations, and the simulation will be carried out in a time-independent radial electric field, which is simply the ambipolar radial electric field in subsection \ref{sec.amEr} and is shown in Fig. \ref{amEr}. Since the magnetic perturbations and kinetic eletron are not included in this simulation, the number of grid points in the $(r,\alpha,\theta,v_{\|},\mu)$ directions is $(450,4,16,64,16)$; the radial grid is uniform; the simulated time step is 10 ion gyro-period. In order to exclude the influence of thermal pressure, we set a uniform temperature $T_0 = 500 eV$ and density profile $N_0 = 3 \times 10^{19} m^{-3}$ here.
\begin{figure}[htbp] 
  \includegraphics[scale=0.6]{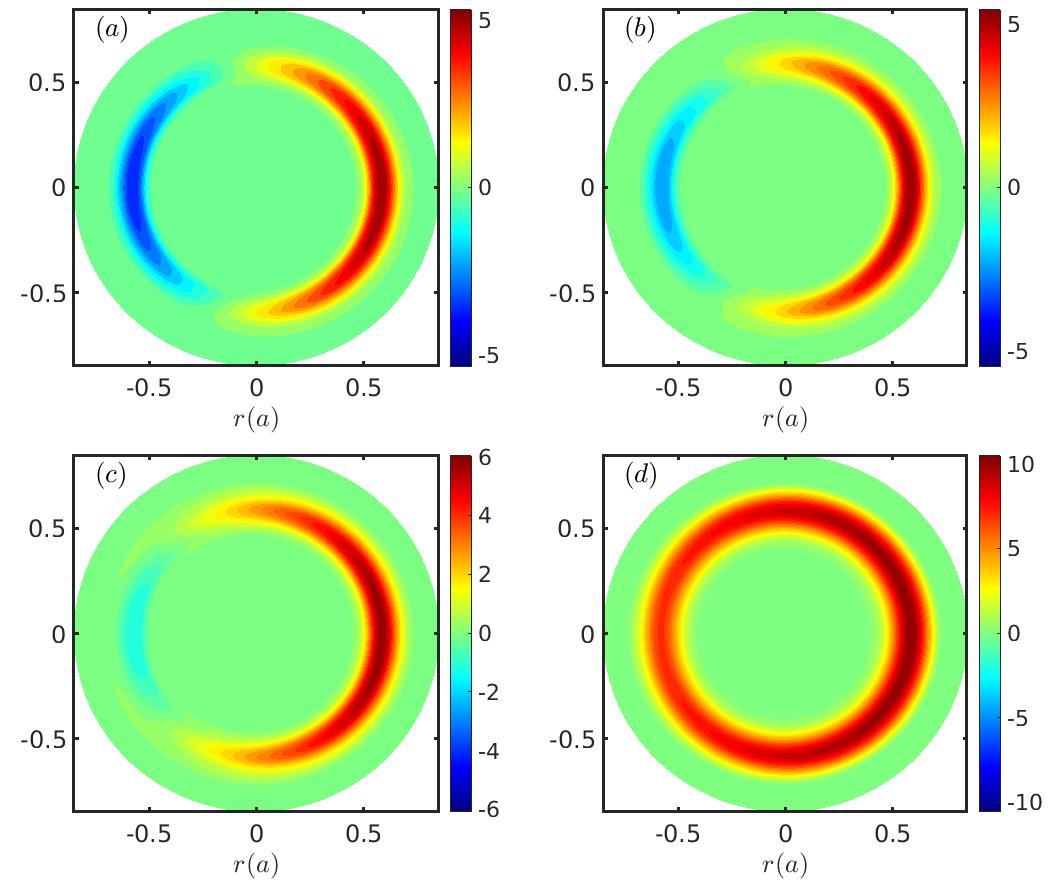}
  \caption{The poloidal structure of $\Gamma_{\zeta}$ at (a) $t_a = 0.18 \tau_{ii}$, (b) $t_b = 0.54 \tau_{ii}$, (c) $t_c = 1.03 \tau_{ii}$, (d) $t_d = 15 \tau_{ii}$.}
  \label{fig.abcd}
\end{figure}

\begin{figure}[htbp] 
  \includegraphics[scale=0.4]{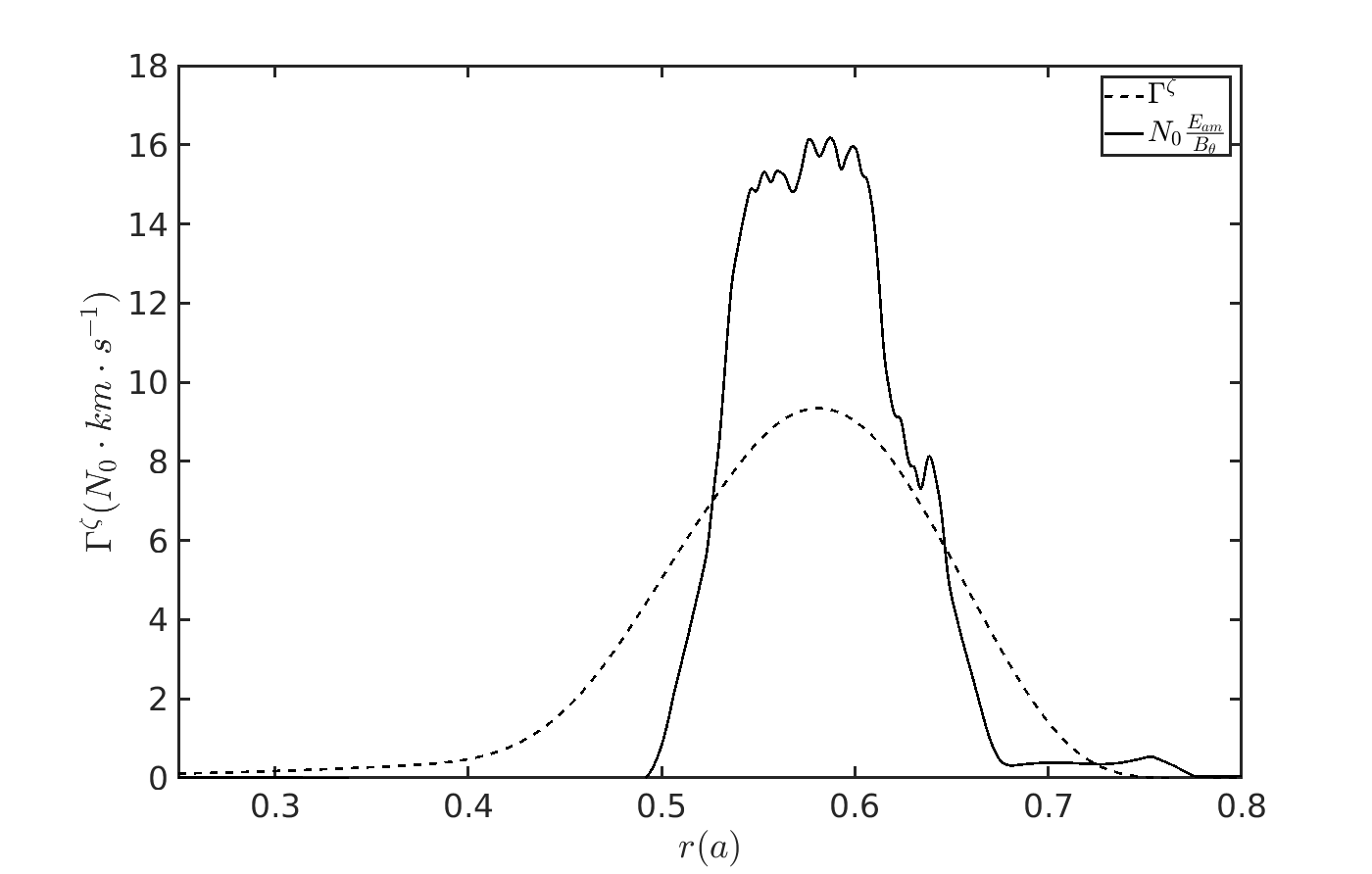}
  \caption{The simulated $\left\langle  \Gamma_{\zeta} \right\rangle_{FA}$ with $T_{0} = 500eV $ compared with the rigid-body rotation theory}
  \label{fig.rigid1}
\end{figure}

   The poloidal structure evolution of $\Gamma_{\zeta}$ is shown in Fig. \ref{fig.abcd}. In this simulation, the return flow is formed on the time scale shorter than an ion-ion collision time, which is consistent with the results discussed in Appendix \ref{Appe.PS-S}. After a few ion-ion collision times, the rigid-body rotation is formed. As shown in Fig. \ref{fig.rigid1}, the radial structure of $\left\langle \Gamma_{\zeta} \right\rangle_{FA}$ agrees roughly with the rigid-body rotation theory. The peak value of $\left\langle \Gamma_{\zeta} \right\rangle_{FA}$ is about $60\%$ of the theoretical value, and $\left\langle \Gamma_{\zeta} \right\rangle_{FA}$ disperses outside the radial electric field region. This is due to the effect of finite orbit width, which also affects the return flow shown in Fig. \ref{CosMt2}(b) in Appendix \ref{Appe.PS-S}. When the ion temperature is lower, the orbit width will be narrower, and the finite orbit width effect will be weakened. Therefore, $\left\langle \Gamma_{\zeta} \right\rangle_{FA}$ will agree better with the rigid-body rotation theory. 

\begin{figure}[htbp] 
  \includegraphics[scale=0.4]{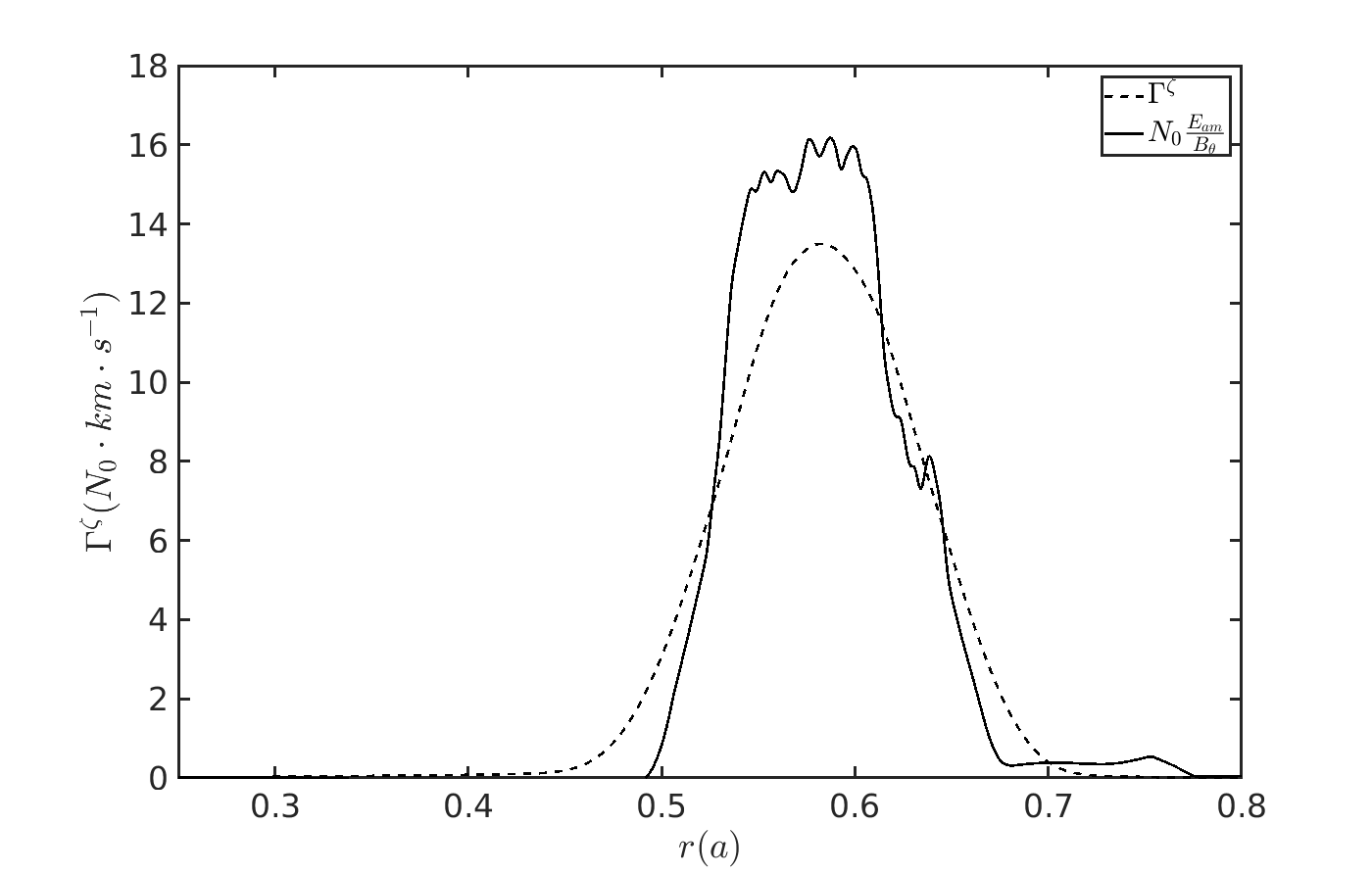}
  \caption{The simulated $\left\langle  \Gamma_{\zeta} \right\rangle_{FA}$ with $T_{0} = 100eV $ compared with the rigid-body rotation theory.}
  \label{fig.rigid2}
\end{figure}

    We set a uniform lower temperature $T_0 = 100 eV$ and density profile $N_0 = 2.68 \times 10^{18} m^{-3}$ in the anther simulation. In this setup, the plasma is still in the banana regime. The orbit width of the $T_0 = 100 eV$ case is $\frac{1}{\sqrt{5}}$ times that of the $T_0 = 500 eV$ case. The simulated radial structure of $\left\langle \Gamma_{\zeta} \right\rangle_{FA}$ is shown in Fig. \ref{fig.rigid2}. The radial structure of $\left\langle \Gamma_{\zeta} \right\rangle_{FA}$ agrees well with the rigid-body rotation theory. The peak value of $\left\langle \Gamma_{\zeta} \right\rangle_{FA}$ is about $85\%$ of the theoretical value, and $\left\langle \Gamma_{\zeta} \right\rangle_{FA}$ is almost within the radial electric field region.


    In Ref. \onlinecite{Schmitz_2019}, the application of RMPs produced a magnetic stochastic layer at the device edge and quickly formed a stable ambipolar radial electric field, which made the radial electric field reverse. Fig. 11 of Ref. \onlinecite{Schmitz_2019} suggests that the plasma toroidal rotation increases correspondingly well after the radial electric field is reversed. Around $\rho = 0.97$, the increment of the radial electric field $\Delta E_r$ is about $5 kV \cdot m^{-1}$, and the increment of the plasma toroidal rotation velocity $\Delta v_{\zeta}$ is about $20 km \cdot s^{-1}$. From the $q = \frac{r B_{\zeta}}{R B_{\theta}}$ profile in Fig. 10 of Ref. \onlinecite{Schmitz_2019}, one can estimate $B_{\theta} = \frac{r B_{\zeta}}{qR} \approx 0.2T$. From thses parameters, we find that $\frac{\Delta E_r}{B_{\theta}} \approx 25 km \cdot s^{-1}$ is in good agreement with the experimentally observed the increment of the plasma toroidal rotation velocity $\Delta v_{\zeta} \approx 20 km \cdot s^{-1}$. This indicates that the effects of RMPs on the toroidal rotation observed in Ref. \onlinecite{Schmitz_2019} can be explained by the ambipolar radial electric field generated by the stochastic magnetic perturbation.

\section{Summary and discussion}

The DIII-D RMP experiment\cite{Schmitz_2019} suggests that the collisional NTV effect is not enough to explain the observed toroidal rotation driven by RMPs. A gyrokinetic simulation of the spontaneous toroidal rotation in a stochastic magnetic field has been carried out, with a self-consistent radial electric field driven by the non-ambipolar radial diffusion due to the magnetic field line braiding. We found that 

  (1) The stochastic magnetic field drives the plasma to toroidally rotate through the ambipolar radial electric field self-consistently formed.

   It is well-known that the $\bm{E}_r\times\bm{B}_0$ flow, with $\bm{E}_r = -\nabla \phi(\psi)$, in a tokamak induces a parallel flow, to make the total flow divergent free,
\begin{align} \label{eq.divfree}
   \nabla \cdot \left( \frac{-\nabla \phi(\psi) \times \bm{B}_0}{B_0^2} + \bm{B}_0 F(\psi,\theta) \right) = 0.
\end{align}

   It was pointed out that with a radial electric field force much stonger than pressure gradient, the solution to Eq. \eqref{eq.divfree} gives the rigid-body torodial rotation as the total flow\cite{Hinton1985}. Following the derivation of the Pfirsch-Schl\"uter current\cite{wesson2011tokamaks}, one finds the parallel return flow from Eq. \eqref{eq.divfree}. In fact, when integrating Eq. \eqref{eq.divfree}, there is a undetermined integral constant.   

Expressing $\bm{B}_0= \nabla \psi \times \nabla \zeta + g(\psi) \nabla \zeta$, from Eq. \eqref{eq.divfree}, one can get
\begin{align} \label{eq.divfree2}
  \partial_\theta F(\psi,\theta) - \partial_\theta \left( \omega(\psi) \frac{g(\psi)}{B^2_0} \right) = 0,
\end{align}
with $\omega(\psi) = - \frac{d\phi}{d\psi}$. The solution of Eq. \eqref{eq.divfree2} is
\begin{align}
  F(\psi,\theta) = \omega(\psi)\frac{g(\psi)}{B^2_0} + C(\psi). 
\end{align}
where $C(\psi)$ is the integral constant with respect to $\theta$. When $C(\psi) = 0$ is chosen, the solution is the toroidal rigid-body flow. When $C(\psi) = - \left\langle \omega(\psi)\frac{g(\psi)}{B^2_0} \right\rangle_{\theta}$ is chosen, with $\left\langle \cdot \right\rangle_{\theta}$ the $\theta$ average operator, the solution is the parallel return flow. In Ref. \onlinecite{Hinton1985}, based on the total flow divergent free, it is directly given that the equilibrium toroidal rotation is a rigid-body rotation, but how the rigid-body rotation is generated and why the return flow is not being the equilibrium toroidal rotation is not explained. According to the above disscussion, there are many solutions to Eq. \eqref{eq.divfree}, in addition to the return flow and rigid-body flow.

  (2) Through our simulation, we found that the spontaneous flow observed in the collisionless simulation(Sec. IIIB-C, Appendix \ref{Appe.PS-S}) is the return flow driven on the time scale of the ion transit time, which is shorter than an ion-ion collision time, by the $\bm{E}_r\times\bm{B}_0$ drift due to the ambipolar radial electric field generated on the electron transit time scale. We have derived the parallel return flow by using the kinetic theory and the characteristic line method in Appendix \ref{Appe.PS}; note that our derivation avoids the issue of integral constant in the usual fluid derivation base on the divergence free condition. 

   Ion-ion collision is considered in the simulation of spontaneous toroidal rotation in a time-independent radial electric field, which is simply the ambipolar radial electric field formed in a stochastic magnetic field.  

  (3) We found that the collisional effect changes the toroidal rotation from the return flow to the rigid-body rotation after a few ion-ion collision times. 

The physics of the rigid-body rotation formed on the time scale of a few ion-ion collision times can be understood as follows. Since the neoclassical ion parallel viscosity damps the poloidal flow\cite{Hirshman1981}, the ultimate flow should be in the toroidal direction, and its perpendicular component should be the $\bm{E}_r \times \bm{B}_0$ flow, therefore, one finds the toroidal rigid-body rotation.  

  (4) Although the simulation results agree well with the theory of the parallel return flow, in the collisionless limit, and the theory of the rigid-body rotation, in the collisional limit, we found that the finite orbit width effect causes the toroidal rotation to diffuse out of the radial electric field region (Appendix \ref{Appe.PS-S} for the return flow; Sec. \ref{sec.rigid} for the rigid-body flow).

  (5) We found that the toroidal rotation observed in the RMP experiment\cite{Schmitz_2019}, can be explained by the rigid-body rotation driven by the ambipolar radial electric field generated by the stochastic magnetic field layer. 

We note that this simulation also provide a test example for the future gyrokinetic simulation on the spontaneous toroidal rotation of plasma associated with the radial electric field.

It should be noted that our simulation is based on a passive transport model, which assumes a given stochastic magnetic perturbation. Note that a stochastic magnetic field layer is reported in the DIII-D RMP experiment\cite{Schmitz_2019}. The point of this paper is that the toroidal rotation is driven through the radial ambipolar electric field. Note that this ambipolar electric field induced by the stochastic magnetic perturbation is insensitive to the perturbation amplitude; in fact, when the perturbation is stochastic, the ambipolar electric field is independent of the perturbation amplitude\cite{Harvey1981,jxyou2023}.  

It should be noted that electron collisional effects, which are important in considering the collisional NTV effect, have been ignored in this work. Note that we have ignored the collisional NTV effect in this work; since the momentum is mainly carried by ions, we have only considered the ion-ion collisional effect in the simulation of relaxation process. Further investigation to consider both the collisional NTV effect and the magnetic filed line braiding effect on the toroidal rotation in the RMP experiment, shall be left for a future work.

\begin{acknowledgments}
  This work was supported by the National Natural Science Foundation
  of China under Grant No. 12075240 and the National MCF Energy R\&D Program of China No. 2019YFE03060000.
\end{acknowledgments}

\appendix

\section{Derivation of the return flow driven by the radial electric field} \label{Appe.PS}

    In this appendix, we use the drift kinetic theory to derive the formula of the return flow driven by a  slowly growing radial electric field $\delta \tilde{\boldsymbol{E}}_r(r,t) = \delta \boldsymbol{E}_r(r) \cdot e^{\gamma t}$ with $\gamma \to 0$. This radial electric field may be generated by the non-ambipolar radial diffusion, such as in simulation results discussed in Sec. \ref{sec.rotaSim} of the main text. 

	The full distribution function $f$ is decomposed into the equilibrium part $F_0$, and the perturbation part $\delta f$. $F_0$  satisfies 
\begin{align} \label{eq.F0}
   \partial_t F_0 + ( v_{\|}\boldsymbol{b}_0 +  \boldsymbol{v}_D) \cdot \nabla F_0 = 0,
\end{align}
where  $\boldsymbol{v}_D$ is the magnetic drift velocity. The solution of Eq. \eqref{eq.F0} can be chosen as a local Maxwellian, $F_0 = F_M$.  $\delta f$ satisfies 
\begin{align} \label{eq.df}
   \partial_t \delta f + ( v_{\|}\boldsymbol{b}_0 +  \boldsymbol{v}_D) \cdot \nabla \delta f = -e_i \boldsymbol{v}_D \cdot \delta \tilde{\boldsymbol{E}}_r \cdot \partial_K F_0,
\end{align}
where $K = \frac12 M_i v_{\|}^2 + \mu B_0$ is the kinetic energy. For thermal ion, $v_D \sim 10^2 m \cdot s^{-1}$, the thermal speed $v_{th} \sim 10^5 m \cdot s^{-1}$, 
\begin{align}
   \frac{ \boldsymbol{v}_D \cdot \nabla \delta f }{ v_{\|}\boldsymbol{b}_0 \cdot \nabla \delta f} \ll 1
\end{align}
hold and one can reduce Eq. \eqref{eq.df} to 
\begin{align} \label{eq.df2}
   \partial_t \delta f +  v_{\|}\boldsymbol{b}_0 \cdot \nabla \delta f = -e_i \boldsymbol{v}_D \cdot \delta \tilde{\boldsymbol{E}}_r \cdot \partial_K F_0.
\end{align}
Note that the magnetic drift term is included in the numerical simulation [see, Eq. \eqref{eq.orbit}, Eq. \eqref{eq.drift}].
In the coordinates $\boldsymbol{Z} = (r,\theta,K,\mu)$, Eq. \eqref{eq.df2} can be expressed as 
\begin{align}\label{eq.df3}
   \partial_t \delta f + \frac{v_{\|}}{qR} \partial_{\theta} \delta f = v_D e_i \delta \tilde{E}_r \sin \theta \partial_K F_0.
\end{align}
Here, $R$ is the major radius. For simplicity, we assume that $v_{\|}$ and $v_d$ are independent of the $\theta$.

Note that $\delta \tilde{E}_r(t\to -\infty)=0$, we shall assume $\delta f(t\to -\infty)=0$. Eq. \eqref{eq.df3} can be solved by using the characteristic line method. The characteristic line is 
\begin{align}
  \tilde{\theta} = \theta + \int_t^{t^{'}} \frac{v_{\|}}{qR} d\tau = \theta + \omega_{\theta} \left( t^{'} - t \right),
\end{align}
with $\omega_{\theta} = \frac{v_{\|}}{qR}$. Then,
\begin{align}
  \delta f &= v_D e_i \delta E_r \partial_K F_0 \int_{-\infty}^t \sin \left( \theta + \omega_{\theta}(t^{'}-t)\right)e^{\gamma t^{'}}dt^{'}   \notag  \\
  &= -v_D e_i \delta E_r \partial_K F_0 \{ \frac{\omega_{\theta}}{\omega_{\theta}^2 + \gamma^2} \cos \theta - \frac{\gamma}{\omega_{\theta}^2 + \gamma^2} \sin \theta \}.
\end{align}
Taking the limit of $\gamma \to 0$, one can obtain
\begin{align}
   \delta f = -v_D e_i \delta E_r \partial_K F_0  \frac{\cos \theta}{\omega_{\theta}} + v_D e_i \delta E_r \partial_K F_0 \delta_D(\omega_{\theta}) \sin \theta ,
\end{align}
with the $\delta_D$ the Dirac-$\delta$ function. Note that the second term is even with respect to $v_{\|}$, it does not contribute to the parallel flow, one can obtain the parallel flow,
\begin{align}  \label{eq.PSE}
  \Gamma_{return} = \int v_{\|} \delta f  d^3\boldsymbol{v} = 2q N_{0i} \frac{\delta E_r}{B_c} \cos \theta.
\end{align}
Obviously, this is a return flow, whose driven mechanism is similar to the familiar Pfirsch-Schl\"uter return flow.


\section{Simulation of the return flow driven by a time-independent radial electric field}\label{Appe.PS-S}
 
In this appendix, we present simulation results, which directly confirm the theory presented in Appendix. \ref{Appe.PS}. 

\begin{figure}[htbp]  
\includegraphics[scale=0.6]{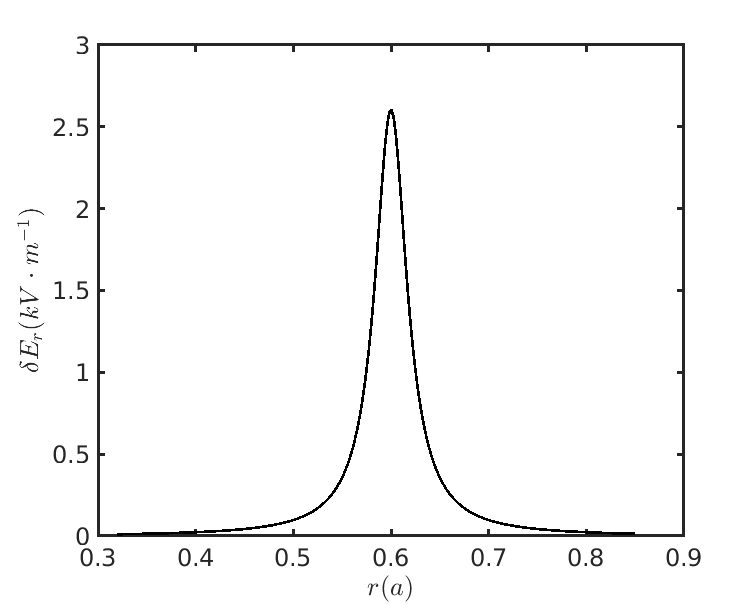}
\caption{The time-independent radial electric field perturbation $\delta E_r$.}  
  \label{Er1}
\end{figure}

\begin{figure}[htbp]  
\includegraphics[scale=0.44]{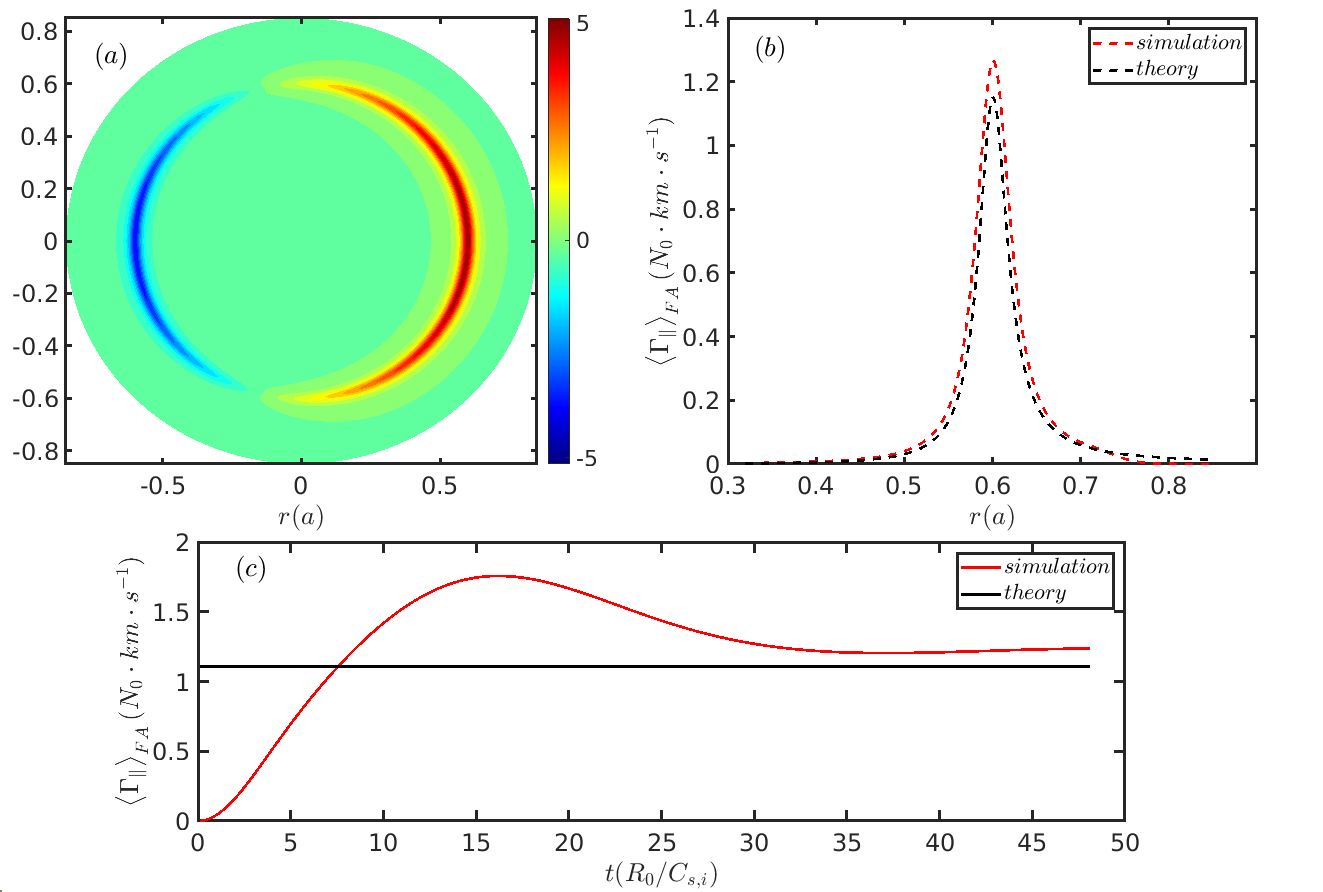}
\caption{ Return flow generated by a time-independent radial electric field. (a) Poloidal distribution of $\Gamma_{\|}(N_{0} \cdot km \cdot s^{-1})$, (b) the simulated $\left\langle  \Gamma_{\|} \right\rangle_{FA}$ compared with theory, and (c) the evolution of $\left\langle  \Gamma_{\|} \right\rangle_{FA}$  at $r = 0.6a$.}  \label{CosMt}
\end{figure}

    Here we do not consider the magnetic perturbations, and simulations are carried out with a time-independent radial electric field perturbation ($\delta \bm{E}_r$) prescribed. Since the magnetic perturbations and kinetic electron are not included in this simulation, the number of grid points in the $(r,\alpha,\theta,v_{\|},\mu)$ directions is $(400,46,16,64,16)$; the radial grid is uniform; the simulated time step is 10 ion gyro-period. Equilibrium parameters are the same as in Sec. \ref{sec.rotaSim}; $M_i = 100M_e$, $T_i \sim 1keV$. The time-independent radial electric field perturbation included in simulation is shown in Fig. \ref{Er1}. The simulation results are shown in Fig. \ref{CosMt}, which agree well with the theory. 

  Fig. \ref{CosMt}(c) shows the time evolution of parallel flow at $r = 0.6a$. As shown in Fig. \ref{CosMt}(c), the parallel return flow rapidly approaches to the theoretical value, with the generation time scale roughly $10\frac{R_0}{C_{s,i}}$($10^{-5}s$), which is much less than an ion-ion collision time($10^{-4}s$). 

\begin{figure}[htbp]  
\includegraphics[scale=0.44]{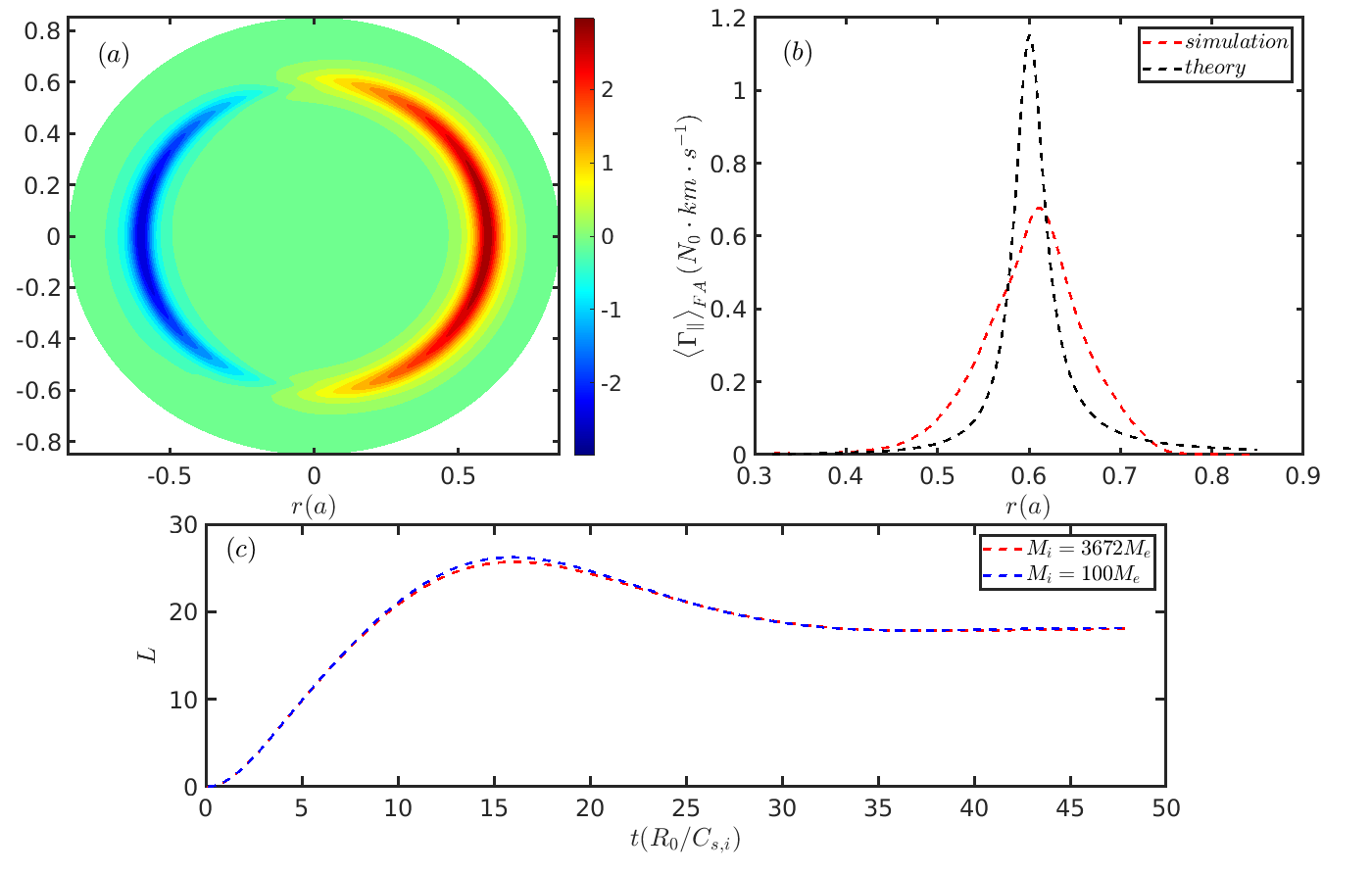}
\caption{Simulation of return flow generated by a time-independent radial electric field, with realistic ion mass. (a) The poloidal distribution of $\Gamma_{\|}(N_{0} \cdot km \cdot s^{-1})$, (b) the simulated radial structure of  $\left\langle  \Gamma_{\|} \right\rangle_{FA}$ compared with the theoretical value at $t = 48R_0/C_{s,i}$, (c) the $L$ evolution of the $M_i=3672M_e$ case and the $M_i = 100M_e$ case. Note that in the time unit, $C_{s,i} = \sqrt{T_i/M_i}$, with $M_i = 100M_e$ and $M_i = 3672 M_e$ for the two cases, respectively.}  \label{CosMt2}
\end{figure}

    We have also carried out a simulation with the realistic ion mass $M_i = 3672M_e$; the results are shown in Fig.\ref{CosMt2}. As shown in Fig. \ref{CosMt2}, the poloidal structure of $\Gamma_{\|}$ is still a cosine structure, but the amplitude of the $\Gamma_{\|}$ is smaller. The amplitude of $\Gamma_{\|}$ is reduced by about half, and the radial structure $\Gamma_{\|}$ is widened. 

The volume integrated parallel momentum can be defined as 
\begin{align}
     L = \int J_{\boldsymbol{X}} \cdot \left\langle  \Gamma_{\|} \right\rangle_{FA} dr.
\end{align}
The $L$ evolution of the $M_i=3672M_e$ case and the $M_i = 100M_e$ case are shown in Fig.\ref{CosMt2}(c). It can be found that the total flow driven by the radial electric field is the same for both cases. In the case of realistic ion mass,$M_i = 3672M_e$, the simulated parallel flow is radially widened about one banana width, comparing to the theoretical results. This is because the return flow formula Eq. \eqref{eq.return} does not take into account of the effects of finite banana width. 

   In the simulation where the electron is a drift-kinetic electron rather than an adiabatic electron, the characteristic time and spatial scales of the ion are much larger than of the electron, which makes the computation very expensive. Replacing $M_i=3672M_e$ with $M_i=100M_e$ can significantly reduce the computational resource consumption. The simulation results of the $M_i=3672M_e$ case and the $M_i=100M_e$ case, suggest that it is reasonable to replace $M_i=3672M_e$ with $M_i=100M_e$.



\bibliography{manuscript}

\begin{thebibliography}{44}%
\makeatletter
\providecommand \@ifxundefined [1]{%
 \@ifx{#1\undefined}
}%
\providecommand \@ifnum [1]{%
 \ifnum #1\expandafter \@firstoftwo
 \else \expandafter \@secondoftwo
 \fi
}%
\providecommand \@ifx [1]{%
 \ifx #1\expandafter \@firstoftwo
 \else \expandafter \@secondoftwo
 \fi
}%
\providecommand \natexlab [1]{#1}%
\providecommand \enquote  [1]{``#1''}%
\providecommand \bibnamefont  [1]{#1}%
\providecommand \bibfnamefont [1]{#1}%
\providecommand \citenamefont [1]{#1}%
\providecommand \href@noop [0]{\@secondoftwo}%
\providecommand \href [0]{\begingroup \@sanitize@url \@href}%
\providecommand \@href[1]{\@@startlink{#1}\@@href}%
\providecommand \@@href[1]{\endgroup#1\@@endlink}%
\providecommand \@sanitize@url [0]{\catcode `\\12\catcode `\$12\catcode `\&12\catcode `\#12\catcode `\^12\catcode `\_12\catcode `\%12\relax}%
\providecommand \@@startlink[1]{}%
\providecommand \@@endlink[0]{}%
\providecommand \url  [0]{\begingroup\@sanitize@url \@url }%
\providecommand \@url [1]{\endgroup\@href {#1}{\urlprefix }}%
\providecommand \urlprefix  [0]{URL }%
\providecommand \Eprint [0]{\href }%
\providecommand \doibase [0]{http://dx.doi.org/}%
\providecommand \selectlanguage [0]{\@gobble}%
\providecommand \bibinfo  [0]{\@secondoftwo}%
\providecommand \bibfield  [0]{\@secondoftwo}%
\providecommand \translation [1]{[#1]}%
\providecommand \BibitemOpen [0]{}%
\providecommand \bibitemStop [0]{}%
\providecommand \bibitemNoStop [0]{.\EOS\space}%
\providecommand \EOS [0]{\spacefactor3000\relax}%
\providecommand \BibitemShut  [1]{\csname bibitem#1\endcsname}%
\let\auto@bib@innerbib\@empty
\bibitem [{\citenamefont {Eriksson}\ \emph {et~al.}(1997)\citenamefont {Eriksson}, \citenamefont {Righi},\ and\ \citenamefont {Zastrow}}]{JET1997}%
  \BibitemOpen
  \bibfield  {author} {\bibinfo {author} {\bibfnamefont {L.-G.}\ \bibnamefont {Eriksson}}, \bibinfo {author} {\bibfnamefont {E.}~\bibnamefont {Righi}}, \ and\ \bibinfo {author} {\bibfnamefont {K.-D.}\ \bibnamefont {Zastrow}},\ }\href@noop {} {\bibfield  {journal} {\bibinfo  {journal} {Plasma Phys. Control. Fusion}\ }\textbf {\bibinfo {volume} {39}},\ \bibinfo {pages} {27} (\bibinfo {year} {1997})}\BibitemShut {NoStop}%
\bibitem [{\citenamefont {Rice}\ \emph {et~al.}(2004)\citenamefont {Rice}, \citenamefont {Lee}, \citenamefont {Marmar}, \citenamefont {Bonoli}, \citenamefont {Granetz}, \citenamefont {Greenwald}, \citenamefont {Hubbard}, \citenamefont {Hutchinson}, \citenamefont {Irby}, \citenamefont {Lin} \emph {et~al.}}]{C-Mod2004}%
  \BibitemOpen
  \bibfield  {author} {\bibinfo {author} {\bibfnamefont {J.}~\bibnamefont {Rice}}, \bibinfo {author} {\bibfnamefont {W.}~\bibnamefont {Lee}}, \bibinfo {author} {\bibfnamefont {E.}~\bibnamefont {Marmar}}, \bibinfo {author} {\bibfnamefont {P.}~\bibnamefont {Bonoli}}, \bibinfo {author} {\bibfnamefont {R.}~\bibnamefont {Granetz}}, \bibinfo {author} {\bibfnamefont {M.}~\bibnamefont {Greenwald}}, \bibinfo {author} {\bibfnamefont {A.}~\bibnamefont {Hubbard}}, \bibinfo {author} {\bibfnamefont {I.}~\bibnamefont {Hutchinson}}, \bibinfo {author} {\bibfnamefont {J.}~\bibnamefont {Irby}}, \bibinfo {author} {\bibfnamefont {Y.}~\bibnamefont {Lin}},  \emph {et~al.},\ }\href@noop {} {\bibfield  {journal} {\bibinfo  {journal} {Nucl. Fusion}\ }\textbf {\bibinfo {volume} {44}},\ \bibinfo {pages} {379} (\bibinfo {year} {2004})}\BibitemShut {NoStop}%
\bibitem [{\citenamefont {Eriksson}\ \emph {et~al.}(2001)\citenamefont {Eriksson}, \citenamefont {Hoang},\ and\ \citenamefont {Bergeaud}}]{Supra2001}%
  \BibitemOpen
  \bibfield  {author} {\bibinfo {author} {\bibfnamefont {L.-G.}\ \bibnamefont {Eriksson}}, \bibinfo {author} {\bibfnamefont {G.}~\bibnamefont {Hoang}}, \ and\ \bibinfo {author} {\bibfnamefont {V.}~\bibnamefont {Bergeaud}},\ }\href@noop {} {\bibfield  {journal} {\bibinfo  {journal} {Nucl. Fusion}\ }\textbf {\bibinfo {volume} {41}},\ \bibinfo {pages} {91} (\bibinfo {year} {2001})}\BibitemShut {NoStop}%
\bibitem [{\citenamefont {deGrassie}\ \emph {et~al.}(2007)\citenamefont {deGrassie}, \citenamefont {Rice}, \citenamefont {Burrell}, \citenamefont {Groebner},\ and\ \citenamefont {Solomon}}]{DIII-D2007}%
  \BibitemOpen
  \bibfield  {author} {\bibinfo {author} {\bibfnamefont {J.~S.}\ \bibnamefont {deGrassie}}, \bibinfo {author} {\bibfnamefont {J.~E.}\ \bibnamefont {Rice}}, \bibinfo {author} {\bibfnamefont {K.~H.}\ \bibnamefont {Burrell}}, \bibinfo {author} {\bibfnamefont {R.~J.}\ \bibnamefont {Groebner}}, \ and\ \bibinfo {author} {\bibfnamefont {W.~M.}\ \bibnamefont {Solomon}},\ }\href@noop {} {\bibfield  {journal} {\bibinfo  {journal} {Phys. Plasmas}\ }\textbf {\bibinfo {volume} {14}},\ \bibinfo {pages} {056115} (\bibinfo {year} {2007})}\BibitemShut {NoStop}%
\bibitem [{\citenamefont {Schmitz}\ \emph {et~al.}(2019)\citenamefont {Schmitz}, \citenamefont {Kriete}, \citenamefont {Wilcox}, \citenamefont {Rhodes}, \citenamefont {Zeng}, \citenamefont {Yan}, \citenamefont {McKee}, \citenamefont {Evans}, \citenamefont {Paz-Soldan}, \citenamefont {Gohil},\ and\ \citenamefont {othes}}]{Schmitz_2019}%
  \BibitemOpen
  \bibfield  {author} {\bibinfo {author} {\bibfnamefont {L.}~\bibnamefont {Schmitz}}, \bibinfo {author} {\bibfnamefont {D.}~\bibnamefont {Kriete}}, \bibinfo {author} {\bibfnamefont {R.}~\bibnamefont {Wilcox}}, \bibinfo {author} {\bibfnamefont {T.}~\bibnamefont {Rhodes}}, \bibinfo {author} {\bibfnamefont {L.}~\bibnamefont {Zeng}}, \bibinfo {author} {\bibfnamefont {Z.}~\bibnamefont {Yan}}, \bibinfo {author} {\bibfnamefont {G.}~\bibnamefont {McKee}}, \bibinfo {author} {\bibfnamefont {T.}~\bibnamefont {Evans}}, \bibinfo {author} {\bibfnamefont {C.}~\bibnamefont {Paz-Soldan}}, \bibinfo {author} {\bibfnamefont {P.}~\bibnamefont {Gohil}}, \ and\ \bibinfo {author} {\bibnamefont {othes}},\ }\href@noop {} {\bibfield  {journal} {\bibinfo  {journal} {Nucl. Fusion}\ }\textbf {\bibinfo {volume} {59}},\ \bibinfo {pages} {126010} (\bibinfo {year} {2019})}\BibitemShut {NoStop}%
\bibitem [{\citenamefont {Seol}\ \emph {et~al.}(2012)\citenamefont {Seol}, \citenamefont {Lee}, \citenamefont {Park}, \citenamefont {Lee}, \citenamefont {Terzolo}, \citenamefont {Shaing}, \citenamefont {You}, \citenamefont {Yun}, \citenamefont {Kim}, \citenamefont {Lee}, \citenamefont {Ko} \emph {et~al.}}]{KSTART2012}%
  \BibitemOpen
  \bibfield  {author} {\bibinfo {author} {\bibfnamefont {J.}~\bibnamefont {Seol}}, \bibinfo {author} {\bibfnamefont {S.~G.}\ \bibnamefont {Lee}}, \bibinfo {author} {\bibfnamefont {B.~H.}\ \bibnamefont {Park}}, \bibinfo {author} {\bibfnamefont {H.~H.}\ \bibnamefont {Lee}}, \bibinfo {author} {\bibfnamefont {L.}~\bibnamefont {Terzolo}}, \bibinfo {author} {\bibfnamefont {K.~C.}\ \bibnamefont {Shaing}}, \bibinfo {author} {\bibfnamefont {K.~I.}\ \bibnamefont {You}}, \bibinfo {author} {\bibfnamefont {G.~S.}\ \bibnamefont {Yun}}, \bibinfo {author} {\bibfnamefont {C.~C.}\ \bibnamefont {Kim}}, \bibinfo {author} {\bibfnamefont {K.~D.}\ \bibnamefont {Lee}}, \bibinfo {author} {\bibfnamefont {W.~H.}\ \bibnamefont {Ko}},  \emph {et~al.},\ }\href@noop {} {\bibfield  {journal} {\bibinfo  {journal} {Phys. Rev. Lett.}\ }\textbf {\bibinfo {volume} {109}},\ \bibinfo {pages} {195003} (\bibinfo {year} {2012})}\BibitemShut {NoStop}%
\bibitem [{\citenamefont {Yang}\ \emph {et~al.}(2019)\citenamefont {Yang}, \citenamefont {Park}, \citenamefont {Na}, \citenamefont {Wang}, \citenamefont {Ko}, \citenamefont {In}, \citenamefont {Lee}, \citenamefont {Lee},\ and\ \citenamefont {Kim}}]{KSTART2019}%
  \BibitemOpen
  \bibfield  {author} {\bibinfo {author} {\bibfnamefont {S.~M.}\ \bibnamefont {Yang}}, \bibinfo {author} {\bibfnamefont {J.-K.}\ \bibnamefont {Park}}, \bibinfo {author} {\bibfnamefont {Y.-S.}\ \bibnamefont {Na}}, \bibinfo {author} {\bibfnamefont {Z.~R.}\ \bibnamefont {Wang}}, \bibinfo {author} {\bibfnamefont {W.~H.}\ \bibnamefont {Ko}}, \bibinfo {author} {\bibfnamefont {Y.}~\bibnamefont {In}}, \bibinfo {author} {\bibfnamefont {J.~H.}\ \bibnamefont {Lee}}, \bibinfo {author} {\bibfnamefont {K.~D.}\ \bibnamefont {Lee}}, \ and\ \bibinfo {author} {\bibfnamefont {S.~K.}\ \bibnamefont {Kim}},\ }\href@noop {} {\bibfield  {journal} {\bibinfo  {journal} {Phys. Rev. Lett.}\ }\textbf {\bibinfo {volume} {123}},\ \bibinfo {pages} {095001} (\bibinfo {year} {2019})}\BibitemShut {NoStop}%
\bibitem [{\citenamefont {Ayten}\ and\ \citenamefont {Westerhof}(2012)}]{NTM2012}%
  \BibitemOpen
  \bibfield  {author} {\bibinfo {author} {\bibfnamefont {B.}~\bibnamefont {Ayten}}\ and\ \bibinfo {author} {\bibfnamefont {E.}~\bibnamefont {Westerhof}},\ }\href@noop {} {\bibfield  {journal} {\bibinfo  {journal} {Phys. Plasmas}\ }\textbf {\bibinfo {volume} {19}},\ \bibinfo {pages} {092506} (\bibinfo {year} {2012})}\BibitemShut {NoStop}%
\bibitem [{\citenamefont {Okabayashi}\ \emph {et~al.}(2002)\citenamefont {Okabayashi}, \citenamefont {Bialek}, \citenamefont {Chance}, \citenamefont {Chu}, \citenamefont {Fredrickson}, \citenamefont {Garofalo}, \citenamefont {Hatcher}, \citenamefont {Jensen}, \citenamefont {Johnson}, \citenamefont {Haye} \emph {et~al.}}]{D3DRWM2002}%
  \BibitemOpen
  \bibfield  {author} {\bibinfo {author} {\bibfnamefont {M.}~\bibnamefont {Okabayashi}}, \bibinfo {author} {\bibfnamefont {J.}~\bibnamefont {Bialek}}, \bibinfo {author} {\bibfnamefont {M.~S.}\ \bibnamefont {Chance}}, \bibinfo {author} {\bibfnamefont {M.~S.}\ \bibnamefont {Chu}}, \bibinfo {author} {\bibfnamefont {E.~D.}\ \bibnamefont {Fredrickson}}, \bibinfo {author} {\bibfnamefont {A.~M.}\ \bibnamefont {Garofalo}}, \bibinfo {author} {\bibfnamefont {R.}~\bibnamefont {Hatcher}}, \bibinfo {author} {\bibfnamefont {T.~H.}\ \bibnamefont {Jensen}}, \bibinfo {author} {\bibfnamefont {L.~C.}\ \bibnamefont {Johnson}}, \bibinfo {author} {\bibfnamefont {R.~J.~L.}\ \bibnamefont {Haye}},  \emph {et~al.},\ }\href@noop {} {\bibfield  {journal} {\bibinfo  {journal} {Plasma Phys. Control. Fusion}\ }\textbf {\bibinfo {volume} {44}},\ \bibinfo {pages} {B339} (\bibinfo {year} {2002})}\BibitemShut {NoStop}%
\bibitem [{\citenamefont {Zheng}\ \emph {et~al.}(2005)\citenamefont {Zheng}, \citenamefont {Kotschenreuther},\ and\ \citenamefont {Chu}}]{Zheng2005}%
  \BibitemOpen
  \bibfield  {author} {\bibinfo {author} {\bibfnamefont {L.-J.}\ \bibnamefont {Zheng}}, \bibinfo {author} {\bibfnamefont {M.}~\bibnamefont {Kotschenreuther}}, \ and\ \bibinfo {author} {\bibfnamefont {M.~S.}\ \bibnamefont {Chu}},\ }\href@noop {} {\bibfield  {journal} {\bibinfo  {journal} {Phys. Rev. Lett.}\ }\textbf {\bibinfo {volume} {95}},\ \bibinfo {pages} {255003} (\bibinfo {year} {2005})}\BibitemShut {NoStop}%
\bibitem [{\citenamefont {Pustovitov}\ and\ \citenamefont {Yanovskiy}(2015)}]{RWM2015}%
  \BibitemOpen
  \bibfield  {author} {\bibinfo {author} {\bibfnamefont {V.~D.}\ \bibnamefont {Pustovitov}}\ and\ \bibinfo {author} {\bibfnamefont {V.~V.}\ \bibnamefont {Yanovskiy}},\ }\href@noop {} {\bibfield  {journal} {\bibinfo  {journal} {Phys. Plasmas}\ }\textbf {\bibinfo {volume} {22}},\ \bibinfo {pages} {032509} (\bibinfo {year} {2015})}\BibitemShut {NoStop}%
\bibitem [{\citenamefont {Li}\ \emph {et~al.}(2022)\citenamefont {Li}, \citenamefont {Liu}, \citenamefont {Liu},\ and\ \citenamefont {Fang}}]{RWM2022}%
  \BibitemOpen
  \bibfield  {author} {\bibinfo {author} {\bibfnamefont {S.}~\bibnamefont {Li}}, \bibinfo {author} {\bibfnamefont {Y.}~\bibnamefont {Liu}}, \bibinfo {author} {\bibfnamefont {C.}~\bibnamefont {Liu}}, \ and\ \bibinfo {author} {\bibfnamefont {Y.}~\bibnamefont {Fang}},\ }\href@noop {} {\bibfield  {journal} {\bibinfo  {journal} {Phys. Plasmas}\ }\textbf {\bibinfo {volume} {29}},\ \bibinfo {pages} {042109} (\bibinfo {year} {2022})}\BibitemShut {NoStop}%
\bibitem [{\citenamefont {Callen}\ \emph {et~al.}(2010)\citenamefont {Callen}, \citenamefont {Hegna},\ and\ \citenamefont {Cole}}]{Callen2010}%
  \BibitemOpen
  \bibfield  {author} {\bibinfo {author} {\bibfnamefont {J.~D.}\ \bibnamefont {Callen}}, \bibinfo {author} {\bibfnamefont {C.~C.}\ \bibnamefont {Hegna}}, \ and\ \bibinfo {author} {\bibfnamefont {A.~J.}\ \bibnamefont {Cole}},\ }\href@noop {} {\bibfield  {journal} {\bibinfo  {journal} {Phys. Plasmas}\ }\textbf {\bibinfo {volume} {17}},\ \bibinfo {pages} {056113} (\bibinfo {year} {2010})}\BibitemShut {NoStop}%
\bibitem [{\citenamefont {Sabbagh}\ \emph {et~al.}(2004)\citenamefont {Sabbagh}, \citenamefont {Bialek}, \citenamefont {Bell}, \citenamefont {Glasser}, \citenamefont {LeBlanc}, \citenamefont {Menard}, \citenamefont {Paoletti}, \citenamefont {Bell}, \citenamefont {Fitzpatrick}, \citenamefont {Fredrickson} \emph {et~al.}}]{NSTX2004}%
  \BibitemOpen
  \bibfield  {author} {\bibinfo {author} {\bibfnamefont {S.}~\bibnamefont {Sabbagh}}, \bibinfo {author} {\bibfnamefont {J.}~\bibnamefont {Bialek}}, \bibinfo {author} {\bibfnamefont {R.}~\bibnamefont {Bell}}, \bibinfo {author} {\bibfnamefont {A.}~\bibnamefont {Glasser}}, \bibinfo {author} {\bibfnamefont {B.}~\bibnamefont {LeBlanc}}, \bibinfo {author} {\bibfnamefont {J.}~\bibnamefont {Menard}}, \bibinfo {author} {\bibfnamefont {F.}~\bibnamefont {Paoletti}}, \bibinfo {author} {\bibfnamefont {M.}~\bibnamefont {Bell}}, \bibinfo {author} {\bibfnamefont {R.}~\bibnamefont {Fitzpatrick}}, \bibinfo {author} {\bibfnamefont {E.}~\bibnamefont {Fredrickson}},  \emph {et~al.},\ }\href@noop {} {\bibfield  {journal} {\bibinfo  {journal} {Nucl. Fusion}\ }\textbf {\bibinfo {volume} {44}},\ \bibinfo {pages} {560} (\bibinfo {year} {2004})}\BibitemShut {NoStop}%
\bibitem [{\citenamefont {Sabbagh}\ \emph {et~al.}(2006)\citenamefont {Sabbagh}, \citenamefont {Sontag}, \citenamefont {Bialek}, \citenamefont {Gates}, \citenamefont {Glasser}, \citenamefont {Menard}, \citenamefont {Zhu}, \citenamefont {Bell}, \citenamefont {Bell}, \citenamefont {Bondeson} \emph {et~al.}}]{NSTX2006}%
  \BibitemOpen
  \bibfield  {author} {\bibinfo {author} {\bibfnamefont {S.}~\bibnamefont {Sabbagh}}, \bibinfo {author} {\bibfnamefont {A.}~\bibnamefont {Sontag}}, \bibinfo {author} {\bibfnamefont {J.}~\bibnamefont {Bialek}}, \bibinfo {author} {\bibfnamefont {D.}~\bibnamefont {Gates}}, \bibinfo {author} {\bibfnamefont {A.}~\bibnamefont {Glasser}}, \bibinfo {author} {\bibfnamefont {J.}~\bibnamefont {Menard}}, \bibinfo {author} {\bibfnamefont {W.}~\bibnamefont {Zhu}}, \bibinfo {author} {\bibfnamefont {M.}~\bibnamefont {Bell}}, \bibinfo {author} {\bibfnamefont {R.}~\bibnamefont {Bell}}, \bibinfo {author} {\bibfnamefont {A.}~\bibnamefont {Bondeson}},  \emph {et~al.},\ }\href@noop {} {\bibfield  {journal} {\bibinfo  {journal} {Nucl. Fusion}\ }\textbf {\bibinfo {volume} {46}},\ \bibinfo {pages} {635} (\bibinfo {year} {2006})}\BibitemShut {NoStop}%
\bibitem [{\citenamefont {Sun}\ \emph {et~al.}(2010{\natexlab{a}})\citenamefont {Sun}, \citenamefont {Liang}, \citenamefont {Koslowski}, \citenamefont {Jachmich}, \citenamefont {Alfier}, \citenamefont {Asunta}, \citenamefont {Corrigan}, \citenamefont {Giroud}, \citenamefont {Gryaznevich}, \citenamefont {Harting} \emph {et~al.}}]{JET2010}%
  \BibitemOpen
  \bibfield  {author} {\bibinfo {author} {\bibfnamefont {Y.}~\bibnamefont {Sun}}, \bibinfo {author} {\bibfnamefont {Y.}~\bibnamefont {Liang}}, \bibinfo {author} {\bibfnamefont {H.~R.}\ \bibnamefont {Koslowski}}, \bibinfo {author} {\bibfnamefont {S.}~\bibnamefont {Jachmich}}, \bibinfo {author} {\bibfnamefont {A.}~\bibnamefont {Alfier}}, \bibinfo {author} {\bibfnamefont {O.}~\bibnamefont {Asunta}}, \bibinfo {author} {\bibfnamefont {G.}~\bibnamefont {Corrigan}}, \bibinfo {author} {\bibfnamefont {C.}~\bibnamefont {Giroud}}, \bibinfo {author} {\bibfnamefont {M.~P.}\ \bibnamefont {Gryaznevich}}, \bibinfo {author} {\bibfnamefont {D.}~\bibnamefont {Harting}},  \emph {et~al.},\ }\href@noop {} {\bibfield  {journal} {\bibinfo  {journal} {Plasma Phys. Control. Fusion}\ }\textbf {\bibinfo {volume} {52}},\ \bibinfo {pages} {105007} (\bibinfo {year} {2010}{\natexlab{a}})}\BibitemShut {NoStop}%
\bibitem [{\citenamefont {Kwak}\ \emph {et~al.}(2013)\citenamefont {Kwak}, \citenamefont {Oh}, \citenamefont {Yang}, \citenamefont {Park}, \citenamefont {Kim}, \citenamefont {Kim}, \citenamefont {Kim}, \citenamefont {Lee}, \citenamefont {Na}, \citenamefont {Kwon} \emph {et~al.}}]{KSTART2013}%
  \BibitemOpen
  \bibfield  {author} {\bibinfo {author} {\bibfnamefont {J.-G.}\ \bibnamefont {Kwak}}, \bibinfo {author} {\bibfnamefont {Y.}~\bibnamefont {Oh}}, \bibinfo {author} {\bibfnamefont {H.}~\bibnamefont {Yang}}, \bibinfo {author} {\bibfnamefont {K.}~\bibnamefont {Park}}, \bibinfo {author} {\bibfnamefont {Y.}~\bibnamefont {Kim}}, \bibinfo {author} {\bibfnamefont {W.}~\bibnamefont {Kim}}, \bibinfo {author} {\bibfnamefont {J.}~\bibnamefont {Kim}}, \bibinfo {author} {\bibfnamefont {S.}~\bibnamefont {Lee}}, \bibinfo {author} {\bibfnamefont {H.}~\bibnamefont {Na}}, \bibinfo {author} {\bibfnamefont {M.}~\bibnamefont {Kwon}},  \emph {et~al.},\ }\href@noop {} {\bibfield  {journal} {\bibinfo  {journal} {Nucl. Fusion}\ }\textbf {\bibinfo {volume} {53}},\ \bibinfo {pages} {104005} (\bibinfo {year} {2013})}\BibitemShut {NoStop}%
\bibitem [{\citenamefont {Shaing}(2003)}]{Shaing2003}%
  \BibitemOpen
  \bibfield  {author} {\bibinfo {author} {\bibfnamefont {K.~C.}\ \bibnamefont {Shaing}},\ }\href@noop {} {\bibfield  {journal} {\bibinfo  {journal} {Phys. Plasmas}\ }\textbf {\bibinfo {volume} {10}},\ \bibinfo {pages} {1443} (\bibinfo {year} {2003})}\BibitemShut {NoStop}%
\bibitem [{\citenamefont {Shaing}\ \emph {et~al.}(2008{\natexlab{a}})\citenamefont {Shaing}, \citenamefont {Cahyna}, \citenamefont {Becoulet}, \citenamefont {Park}, \citenamefont {Sabbagh},\ and\ \citenamefont {Chu}}]{Shaing2008}%
  \BibitemOpen
  \bibfield  {author} {\bibinfo {author} {\bibfnamefont {K.~C.}\ \bibnamefont {Shaing}}, \bibinfo {author} {\bibfnamefont {P.}~\bibnamefont {Cahyna}}, \bibinfo {author} {\bibfnamefont {M.}~\bibnamefont {Becoulet}}, \bibinfo {author} {\bibfnamefont {J.-K.}\ \bibnamefont {Park}}, \bibinfo {author} {\bibfnamefont {S.~A.}\ \bibnamefont {Sabbagh}}, \ and\ \bibinfo {author} {\bibfnamefont {M.~S.}\ \bibnamefont {Chu}},\ }\href@noop {} {\bibfield  {journal} {\bibinfo  {journal} {Phys. Plasmas}\ }\textbf {\bibinfo {volume} {15}},\ \bibinfo {pages} {082506} (\bibinfo {year} {2008}{\natexlab{a}})}\BibitemShut {NoStop}%
\bibitem [{\citenamefont {Shaing}\ \emph {et~al.}(2008{\natexlab{b}})\citenamefont {Shaing}, \citenamefont {Sabbagh},\ and\ \citenamefont {Chu}}]{Shaing2009a}%
  \BibitemOpen
  \bibfield  {author} {\bibinfo {author} {\bibfnamefont {K.~C.}\ \bibnamefont {Shaing}}, \bibinfo {author} {\bibfnamefont {S.~A.}\ \bibnamefont {Sabbagh}}, \ and\ \bibinfo {author} {\bibfnamefont {M.~S.}\ \bibnamefont {Chu}},\ }\href@noop {} {\bibfield  {journal} {\bibinfo  {journal} {Plasma Phys. Control. Fusion}\ }\textbf {\bibinfo {volume} {51}},\ \bibinfo {pages} {035004} (\bibinfo {year} {2008}{\natexlab{b}})}\BibitemShut {NoStop}%
\bibitem [{\citenamefont {Shaing}\ \emph {et~al.}(2008{\natexlab{c}})\citenamefont {Shaing}, \citenamefont {Sabbagh},\ and\ \citenamefont {Chu}}]{Shaing2009b}%
  \BibitemOpen
  \bibfield  {author} {\bibinfo {author} {\bibfnamefont {K.~C.}\ \bibnamefont {Shaing}}, \bibinfo {author} {\bibfnamefont {S.~A.}\ \bibnamefont {Sabbagh}}, \ and\ \bibinfo {author} {\bibfnamefont {M.~S.}\ \bibnamefont {Chu}},\ }\href@noop {} {\bibfield  {journal} {\bibinfo  {journal} {Plasma Phys. Control. Fusion}\ }\textbf {\bibinfo {volume} {51}},\ \bibinfo {pages} {035009} (\bibinfo {year} {2008}{\natexlab{c}})}\BibitemShut {NoStop}%
\bibitem [{\citenamefont {Shaing}\ \emph {et~al.}(2009{\natexlab{a}})\citenamefont {Shaing}, \citenamefont {Sabbagh},\ and\ \citenamefont {Chu}}]{Shaing2009c}%
  \BibitemOpen
  \bibfield  {author} {\bibinfo {author} {\bibfnamefont {K.~C.}\ \bibnamefont {Shaing}}, \bibinfo {author} {\bibfnamefont {S.~A.}\ \bibnamefont {Sabbagh}}, \ and\ \bibinfo {author} {\bibfnamefont {M.~S.}\ \bibnamefont {Chu}},\ }\href@noop {} {\bibfield  {journal} {\bibinfo  {journal} {Plasma Phys. Control. Fusion}\ }\textbf {\bibinfo {volume} {51}},\ \bibinfo {pages} {055003} (\bibinfo {year} {2009}{\natexlab{a}})}\BibitemShut {NoStop}%
\bibitem [{\citenamefont {Shaing}\ \emph {et~al.}(2009{\natexlab{b}})\citenamefont {Shaing}, \citenamefont {Chu},\ and\ \citenamefont {Sabbagh}}]{Shaing2009d}%
  \BibitemOpen
  \bibfield  {author} {\bibinfo {author} {\bibfnamefont {K.~C.}\ \bibnamefont {Shaing}}, \bibinfo {author} {\bibfnamefont {M.~S.}\ \bibnamefont {Chu}}, \ and\ \bibinfo {author} {\bibfnamefont {S.~A.}\ \bibnamefont {Sabbagh}},\ }\href@noop {} {\bibfield  {journal} {\bibinfo  {journal} {Plasma Phys. Control. Fusion}\ }\textbf {\bibinfo {volume} {51}},\ \bibinfo {pages} {075015} (\bibinfo {year} {2009}{\natexlab{b}})}\BibitemShut {NoStop}%
\bibitem [{\citenamefont {Park}\ \emph {et~al.}(2009)\citenamefont {Park}, \citenamefont {Boozer},\ and\ \citenamefont {Menard}}]{JK-Park2009}%
  \BibitemOpen
  \bibfield  {author} {\bibinfo {author} {\bibfnamefont {J.-K.}\ \bibnamefont {Park}}, \bibinfo {author} {\bibfnamefont {A.~H.}\ \bibnamefont {Boozer}}, \ and\ \bibinfo {author} {\bibfnamefont {J.~E.}\ \bibnamefont {Menard}},\ }\href@noop {} {\bibfield  {journal} {\bibinfo  {journal} {Phys. Rev. Lett.}\ }\textbf {\bibinfo {volume} {102}},\ \bibinfo {pages} {065002} (\bibinfo {year} {2009})}\BibitemShut {NoStop}%
\bibitem [{\citenamefont {Park}(2011)}]{JK-Park2011}%
  \BibitemOpen
  \bibfield  {author} {\bibinfo {author} {\bibfnamefont {J.-K.}\ \bibnamefont {Park}},\ }\href@noop {} {\bibfield  {journal} {\bibinfo  {journal} {Phys. Plasmas}\ }\textbf {\bibinfo {volume} {18}},\ \bibinfo {pages} {110702} (\bibinfo {year} {2011})}\BibitemShut {NoStop}%
\bibitem [{\citenamefont {Sun}\ \emph {et~al.}(2010{\natexlab{b}})\citenamefont {Sun}, \citenamefont {Liang}, \citenamefont {Shaing}, \citenamefont {Koslowski}, \citenamefont {Wiegmann},\ and\ \citenamefont {Zhang}}]{Sun2010}%
  \BibitemOpen
  \bibfield  {author} {\bibinfo {author} {\bibfnamefont {Y.}~\bibnamefont {Sun}}, \bibinfo {author} {\bibfnamefont {Y.}~\bibnamefont {Liang}}, \bibinfo {author} {\bibfnamefont {K.~C.}\ \bibnamefont {Shaing}}, \bibinfo {author} {\bibfnamefont {H.~R.}\ \bibnamefont {Koslowski}}, \bibinfo {author} {\bibfnamefont {C.}~\bibnamefont {Wiegmann}}, \ and\ \bibinfo {author} {\bibfnamefont {T.}~\bibnamefont {Zhang}},\ }\href@noop {} {\bibfield  {journal} {\bibinfo  {journal} {Phys. Rev. Lett.}\ }\textbf {\bibinfo {volume} {105}},\ \bibinfo {pages} {145002} (\bibinfo {year} {2010}{\natexlab{b}})}\BibitemShut {NoStop}%
\bibitem [{\citenamefont {Sun}\ \emph {et~al.}(2011)\citenamefont {Sun}, \citenamefont {Liang}, \citenamefont {Shaing}, \citenamefont {Koslowski}, \citenamefont {Wiegmann},\ and\ \citenamefont {Zhang}}]{Sun2011}%
  \BibitemOpen
  \bibfield  {author} {\bibinfo {author} {\bibfnamefont {Y.}~\bibnamefont {Sun}}, \bibinfo {author} {\bibfnamefont {Y.}~\bibnamefont {Liang}}, \bibinfo {author} {\bibfnamefont {K.}~\bibnamefont {Shaing}}, \bibinfo {author} {\bibfnamefont {H.}~\bibnamefont {Koslowski}}, \bibinfo {author} {\bibfnamefont {C.}~\bibnamefont {Wiegmann}}, \ and\ \bibinfo {author} {\bibfnamefont {T.}~\bibnamefont {Zhang}},\ }\href@noop {} {\bibfield  {journal} {\bibinfo  {journal} {Nucl. Fusion}\ }\textbf {\bibinfo {volume} {51}},\ \bibinfo {pages} {053015} (\bibinfo {year} {2011})}\BibitemShut {NoStop}%
\bibitem [{\citenamefont {Sun}\ \emph {et~al.}(2013)\citenamefont {Sun}, \citenamefont {Shaing}, \citenamefont {Liang}, \citenamefont {Shen},\ and\ \citenamefont {Wan}}]{Sun2013}%
  \BibitemOpen
  \bibfield  {author} {\bibinfo {author} {\bibfnamefont {Y.}~\bibnamefont {Sun}}, \bibinfo {author} {\bibfnamefont {K.}~\bibnamefont {Shaing}}, \bibinfo {author} {\bibfnamefont {Y.}~\bibnamefont {Liang}}, \bibinfo {author} {\bibfnamefont {B.}~\bibnamefont {Shen}}, \ and\ \bibinfo {author} {\bibfnamefont {B.}~\bibnamefont {Wan}},\ }\href@noop {} {\bibfield  {journal} {\bibinfo  {journal} {Nucl. Fusion}\ }\textbf {\bibinfo {volume} {53}},\ \bibinfo {pages} {073026} (\bibinfo {year} {2013})}\BibitemShut {NoStop}%
\bibitem [{\citenamefont {Liu}\ \emph {et~al.}(2013)\citenamefont {Liu}, \citenamefont {Kirk},\ and\ \citenamefont {Sun}}]{Liu2013}%
  \BibitemOpen
  \bibfield  {author} {\bibinfo {author} {\bibfnamefont {Y.}~\bibnamefont {Liu}}, \bibinfo {author} {\bibfnamefont {A.}~\bibnamefont {Kirk}}, \ and\ \bibinfo {author} {\bibfnamefont {Y.}~\bibnamefont {Sun}},\ }\href@noop {} {\bibfield  {journal} {\bibinfo  {journal} {Phys. Plasmas}\ }\textbf {\bibinfo {volume} {20}},\ \bibinfo {pages} {042503} (\bibinfo {year} {2013})}\BibitemShut {NoStop}%
\bibitem [{\citenamefont {Sun}\ \emph {et~al.}(2019)\citenamefont {Sun}, \citenamefont {Li}, \citenamefont {He},\ and\ \citenamefont {Shaing}}]{Sun2019}%
  \BibitemOpen
  \bibfield  {author} {\bibinfo {author} {\bibfnamefont {Y.}~\bibnamefont {Sun}}, \bibinfo {author} {\bibfnamefont {X.}~\bibnamefont {Li}}, \bibinfo {author} {\bibfnamefont {K.}~\bibnamefont {He}}, \ and\ \bibinfo {author} {\bibfnamefont {K.~C.}\ \bibnamefont {Shaing}},\ }\href@noop {} {\bibfield  {journal} {\bibinfo  {journal} {Phys. Plasmas}\ }\textbf {\bibinfo {volume} {26}},\ \bibinfo {pages} {072504} (\bibinfo {year} {2019})}\BibitemShut {NoStop}%
\bibitem [{\citenamefont {Li}\ \emph {et~al.}(2019)\citenamefont {Li}, \citenamefont {Sun}, \citenamefont {Wang}, \citenamefont {Zang}, \citenamefont {Li}, \citenamefont {Liu}, \citenamefont {Shi}, \citenamefont {Li}, \citenamefont {Hao}, \citenamefont {Gu},\ and\ \citenamefont {Shen}}]{EAST2019}%
  \BibitemOpen
  \bibfield  {author} {\bibinfo {author} {\bibfnamefont {X.~Y.}\ \bibnamefont {Li}}, \bibinfo {author} {\bibfnamefont {Y.~W.}\ \bibnamefont {Sun}}, \bibinfo {author} {\bibfnamefont {H.~H.}\ \bibnamefont {Wang}}, \bibinfo {author} {\bibfnamefont {Q.}~\bibnamefont {Zang}}, \bibinfo {author} {\bibfnamefont {Y.~Y.}\ \bibnamefont {Li}}, \bibinfo {author} {\bibfnamefont {H.~Q.}\ \bibnamefont {Liu}}, \bibinfo {author} {\bibfnamefont {T.~H.}\ \bibnamefont {Shi}}, \bibinfo {author} {\bibfnamefont {G.~Q.}\ \bibnamefont {Li}}, \bibinfo {author} {\bibfnamefont {B.~L.}\ \bibnamefont {Hao}}, \bibinfo {author} {\bibfnamefont {S.}~\bibnamefont {Gu}}, \ and\ \bibinfo {author} {\bibfnamefont {B.}~\bibnamefont {Shen}},\ }\href@noop {} {\bibfield  {journal} {\bibinfo  {journal} {Phys. Plasmas}\ }\textbf {\bibinfo {volume} {26}},\ \bibinfo {pages} {052512} (\bibinfo {year} {2019})}\BibitemShut {NoStop}%
\bibitem [{\citenamefont {Liu}\ \emph {et~al.}(2010)\citenamefont {Liu}, \citenamefont {Kirk},\ and\ \citenamefont {Nardon}}]{Liu2010}%
  \BibitemOpen
  \bibfield  {author} {\bibinfo {author} {\bibfnamefont {Y.}~\bibnamefont {Liu}}, \bibinfo {author} {\bibfnamefont {A.}~\bibnamefont {Kirk}}, \ and\ \bibinfo {author} {\bibfnamefont {E.}~\bibnamefont {Nardon}},\ }\href@noop {} {\bibfield  {journal} {\bibinfo  {journal} {Phys. Plasmas}\ }\textbf {\bibinfo {volume} {17}},\ \bibinfo {pages} {122502} (\bibinfo {year} {2010})}\BibitemShut {NoStop}%
\bibitem [{\citenamefont {Rechester}\ and\ \citenamefont {Rosenbluth}(1978)}]{RR1978}%
  \BibitemOpen
  \bibfield  {author} {\bibinfo {author} {\bibfnamefont {A.~B.}\ \bibnamefont {Rechester}}\ and\ \bibinfo {author} {\bibfnamefont {M.~N.}\ \bibnamefont {Rosenbluth}},\ }\href@noop {} {\bibfield  {journal} {\bibinfo  {journal} {Phys. Rev. Lett.}\ }\textbf {\bibinfo {volume} {40}} (\bibinfo {year} {1978})}\BibitemShut {NoStop}%
\bibitem [{\citenamefont {Harvey}\ \emph {et~al.}(1981)\citenamefont {Harvey}, \citenamefont {McCoy}, \citenamefont {Hsu},\ and\ \citenamefont {Mirin}}]{Harvey1981}%
  \BibitemOpen
  \bibfield  {author} {\bibinfo {author} {\bibfnamefont {R.~W.}\ \bibnamefont {Harvey}}, \bibinfo {author} {\bibfnamefont {M.~G.}\ \bibnamefont {McCoy}}, \bibinfo {author} {\bibfnamefont {J.~Y.}\ \bibnamefont {Hsu}}, \ and\ \bibinfo {author} {\bibfnamefont {A.~A.}\ \bibnamefont {Mirin}},\ }\href@noop {} {\bibfield  {journal} {\bibinfo  {journal} {Phys. Rev. Lett.}\ }\textbf {\bibinfo {volume} {47}},\ \bibinfo {pages} {102} (\bibinfo {year} {1981})}\BibitemShut {NoStop}%
\bibitem [{\citenamefont {You}\ and\ \citenamefont {Wang}(2023)}]{jxyou2023}%
  \BibitemOpen
  \bibfield  {author} {\bibinfo {author} {\bibfnamefont {J.}~\bibnamefont {You}}\ and\ \bibinfo {author} {\bibfnamefont {S.}~\bibnamefont {Wang}},\ }\href@noop {} {\bibfield  {journal} {\bibinfo  {journal} {AIP Advances}\ }\textbf {\bibinfo {volume} {13}},\ \bibinfo {pages} {095114} (\bibinfo {year} {2023})}\BibitemShut {NoStop}%
\bibitem [{\citenamefont {Ye}\ \emph {et~al.}(2016)\citenamefont {Ye}, \citenamefont {Xu}, \citenamefont {Xiao}, \citenamefont {Dai},\ and\ \citenamefont {Wang}}]{Ye2016}%
  \BibitemOpen
  \bibfield  {author} {\bibinfo {author} {\bibfnamefont {L.}~\bibnamefont {Ye}}, \bibinfo {author} {\bibfnamefont {Y.}~\bibnamefont {Xu}}, \bibinfo {author} {\bibfnamefont {X.}~\bibnamefont {Xiao}}, \bibinfo {author} {\bibfnamefont {Z.}~\bibnamefont {Dai}}, \ and\ \bibinfo {author} {\bibfnamefont {S.}~\bibnamefont {Wang}},\ }\href@noop {} {\bibfield  {journal} {\bibinfo  {journal} {J. Comput. Phys.}\ }\textbf {\bibinfo {volume} {316}},\ \bibinfo {pages} {180} (\bibinfo {year} {2016})}\BibitemShut {NoStop}%
\bibitem [{\citenamefont {Xu}\ \emph {et~al.}(2017)\citenamefont {Xu}, \citenamefont {Ye}, \citenamefont {Dai}, \citenamefont {Xiao},\ and\ \citenamefont {Wang}}]{Xu2017}%
  \BibitemOpen
  \bibfield  {author} {\bibinfo {author} {\bibfnamefont {Y.}~\bibnamefont {Xu}}, \bibinfo {author} {\bibfnamefont {L.}~\bibnamefont {Ye}}, \bibinfo {author} {\bibfnamefont {Z.}~\bibnamefont {Dai}}, \bibinfo {author} {\bibfnamefont {X.}~\bibnamefont {Xiao}}, \ and\ \bibinfo {author} {\bibfnamefont {S.}~\bibnamefont {Wang}},\ }\href@noop {} {\bibfield  {journal} {\bibinfo  {journal} {Phys. Plasmas}\ }\textbf {\bibinfo {volume} {24}},\ \bibinfo {pages} {082515} (\bibinfo {year} {2017})}\BibitemShut {NoStop}%
\bibitem [{\citenamefont {Wang}(2012)}]{Wang12}%
  \BibitemOpen
  \bibfield  {author} {\bibinfo {author} {\bibfnamefont {S.}~\bibnamefont {Wang}},\ }\href@noop {} {\bibfield  {journal} {\bibinfo  {journal} {Phys. Plasmas}\ }\textbf {\bibinfo {volume} {19}},\ \bibinfo {pages} {062504} (\bibinfo {year} {2012})}\BibitemShut {NoStop}%
\bibitem [{\citenamefont {Wang}(2013)}]{Wang13}%
  \BibitemOpen
  \bibfield  {author} {\bibinfo {author} {\bibfnamefont {S.}~\bibnamefont {Wang}},\ }\href@noop {} {\bibfield  {journal} {\bibinfo  {journal} {Phys. Plasmas}\ }\textbf {\bibinfo {volume} {20}},\ \bibinfo {pages} {082312} (\bibinfo {year} {2013})}\BibitemShut {NoStop}%
\bibitem [{\citenamefont {Lee}(1983)}]{WWLEE1983}%
  \BibitemOpen
  \bibfield  {author} {\bibinfo {author} {\bibfnamefont {W.~W.}\ \bibnamefont {Lee}},\ }\href@noop {} {\bibfield  {journal} {\bibinfo  {journal} {Phys. Fluids}\ }\textbf {\bibinfo {volume} {26}},\ \bibinfo {pages} {556} (\bibinfo {year} {1983})}\BibitemShut {NoStop}%
\bibitem [{\citenamefont {Park}\ \emph {et~al.}(2010)\citenamefont {Park}, \citenamefont {Chang}, \citenamefont {Joseph},\ and\ \citenamefont {Moyer}}]{Park2010}%
  \BibitemOpen
  \bibfield  {author} {\bibinfo {author} {\bibfnamefont {G.}~\bibnamefont {Park}}, \bibinfo {author} {\bibfnamefont {C.~S.}\ \bibnamefont {Chang}}, \bibinfo {author} {\bibfnamefont {I.}~\bibnamefont {Joseph}}, \ and\ \bibinfo {author} {\bibfnamefont {R.~A.}\ \bibnamefont {Moyer}},\ }\href@noop {} {\bibfield  {journal} {\bibinfo  {journal} {Phys. Plasmas}\ }\textbf {\bibinfo {volume} {17}},\ \bibinfo {pages} {102503} (\bibinfo {year} {2010})}\BibitemShut {NoStop}%
\bibitem [{\citenamefont {Wesson}(2011)}]{wesson2011tokamaks}%
  \BibitemOpen
  \bibfield  {author} {\bibinfo {author} {\bibfnamefont {J.}~\bibnamefont {Wesson}},\ }\href@noop {} {\emph {\bibinfo {title} {$Tokamaks$}}},\ \bibinfo {edition} {4th}\ ed.\ (\bibinfo  {publisher} {Clarendon Press},\ \bibinfo {address} {Oxford},\ \bibinfo {year} {2011})\ pp.\ \bibinfo {pages} {149--152}\BibitemShut {NoStop}%
\bibitem [{\citenamefont {Hinton}\ and\ \citenamefont {Wong}(1985)}]{Hinton1985}%
  \BibitemOpen
  \bibfield  {author} {\bibinfo {author} {\bibfnamefont {F.~L.}\ \bibnamefont {Hinton}}\ and\ \bibinfo {author} {\bibfnamefont {S.~K.}\ \bibnamefont {Wong}},\ }\href@noop {} {\bibfield  {journal} {\bibinfo  {journal} {Phys. Fluids}\ }\textbf {\bibinfo {volume} {28}},\ \bibinfo {pages} {3082} (\bibinfo {year} {1985})}\BibitemShut {NoStop}%
\bibitem [{\citenamefont {Hirshman}\ and\ \citenamefont {Sigmar}(1981)}]{Hirshman1981}%
  \BibitemOpen
  \bibfield  {author} {\bibinfo {author} {\bibfnamefont {S.}~\bibnamefont {Hirshman}}\ and\ \bibinfo {author} {\bibfnamefont {D.}~\bibnamefont {Sigmar}},\ }\href@noop {} {\bibfield  {journal} {\bibinfo  {journal} {Nucl. Fusion}\ }\textbf {\bibinfo {volume} {21}},\ \bibinfo {pages} {1079} (\bibinfo {year} {1981})}\BibitemShut {NoStop}%
\end{thebibliography}%

\end{document}